\documentclass[final,twocolumn]{IEEEtran}
\usepackage{amssymb,amsmath,float,caption,tikz,booktabs,cite,paralist,epstopdf,setspace,steinmetz,pgfplots,subcaption,soul}
\usetikzlibrary{patterns}

\long\def\comment#1{}

\newfont{\bbb}{msbm10 scaled 700}

\newfont{\bb}{msbm10 scaled 1100}

\newcommand{\EE}{\mbox{\bb E}}

\newcommand{\dv}{{\bf d}}

\newcommand{\hv}{{\bf h}}

\newcommand{\xv}{{\bf x}}
\newcommand{\yv}{{\bf y}}
\newcommand{\zv}{{\bf z}}
\newcommand{\zerov}{{\bf 0}}

\newcommand{\Am}{{\bf A}}

\newcommand{\Cm}{{\bf C}}

\newcommand{\Em}{{\bf E}}
\newcommand{\Fm}{{\bf F}}

\newcommand{\Hm}{{\bf H}}
\newcommand{\Id}{{\bf I}}

\newcommand{\Mm}{{\bf M}}

\newcommand{\Pm}{{\bf P}}

\newcommand{\Rm}{{\bf R}}
\newcommand{\Sm}{{\bf S}}

\newcommand{\Xm}{{\bf X}}

\newcommand{\Ac}{{\cal A}}

\newcommand{\Cc}{{\cal C}}

\newcommand{\Ic}{{\cal I}}

\newcommand{\Nc}{{\cal N}}
\newcommand{\Oc}{{\cal O}}

\newcommand{\Rc}{{\cal R}}
\newcommand{\Sc}{{\cal S}}
\newcommand{\Tc}{{\cal T}}

\newcommand{\Lambdam}{\hbox{\boldmath$\Lambda$}}

\newcommand{\Psim}{\hbox{\boldmath$\Psi$}}

\newcommand{\diag}{{\hbox{diag}}}

\newcommand{\herm}{{\sf H}}

\newcommand{\Ifd}{\mbox{$\boldsymbol{\mathcal{I}}$}}

\newcommand{\Xfd}{\mbox{$\boldsymbol{\mathcal{X}}$}}
\newcommand{\Yfd}{\mbox{$\boldsymbol{\mathcal{Y}}$}}
\newcommand{\Zfd}{\mbox{$\boldsymbol{\mathcal{Z}}$}}

\usepackage{algorithmicx, algpseudocode}
\begin{document}
\title{Narrowband Interference Mitigation in SC-FDMA Using Bayesian Sparse Recovery}
\author{Anum Ali, Mudassir Masood, Muhammad S. Sohail, Samir Al-Ghadhban and Tareq Y. Al-Naffouri
\thanks{This work was supported in part by King Fahd University of Petroleum and Minerals (KFUPM) and King Abdullah University of Science and Technology (KAUST) through project number EE002355, and in part by King Abdulaziz City for Science and Technology (KACST) through the Science \& Technology Unit at KFUPM through project number 09-ELE781-4 as part of the National Science, Technology and Innovation Plan.}
\thanks{A. Ali, M. Masood and T. Y. Al-Naffouri are with the department of Electrical Engineering, KAUST, Thuwal, Saudi Arabia \mbox{(e-mail: \{anum.ali,mudassir.masood,tareq.alnaffouri\}@kaust.edu.sa)}. S. Al-Ghadhban is with the department of Electrical Engineering, KFUPM, Dhahran, Saudi Arabia \mbox{(e-mail: samir@kfupm.edu.sa)}.  M. S. Sohail is with the Department of Electronic and Computer Engineering, Hong Kong University of Science and Technology, Hong Kong \mbox{(email: mssohail@ust.hk)}.}
\thanks{T. Y. Al-Naffouri is also associated with the department of Electrical Engineering, KFUPM, Dhahran, Saudi Arabia.}%
}
\maketitle
\begin{abstract}
This paper presents a novel narrowband interference (NBI) mitigation scheme for SC-FDMA systems. The proposed NBI cancellation scheme exploits the frequency domain sparsity of the unknown signal and adopts a low complexity Bayesian sparse recovery procedure. At the transmitter, a few randomly chosen sub-carriers are kept data free to sense the NBI signal at the receiver. Further, it is noted that in practice, the sparsity of the NBI signal is destroyed by a grid mismatch between NBI sources and the system under consideration. Towards this end, first an accurate grid mismatch model is presented that is capable of assuming independent offsets for multiple NBI sources. Secondly, prior to NBI reconstruction,  the sparsity of the unknown signal is restored by employing a sparsifying transform. To improve the spectral efficiency of the proposed scheme, a data-aided NBI recovery procedure is outlined that relies on adaptively selecting a subset of data carriers and uses them as additional measurements to enhance the NBI estimation. Finally, the proposed scheme is extended to single-input multi-output systems by performing a collaborative NBI support search over all antennas. Numerical results are presented that depict the suitability of the proposed scheme for NBI mitigation.
\end{abstract}
\section{Introduction}\label{secIntro}
\IEEEPARstart{O}{rthogonal} frequency division multiple access (OFDMA) has been used extensively for uplink communications due to its robustness against multipath fading and simple equalization \cite{zhu2012chunk}. However, the transmission signal in OFDMA is the sum of orthogonal sinusoids (with random amplitudes and phases), causing high peak-to-average power ratio (PAPR). The conflicting interest between linearity and power efficiency of the power amplifier renders the high PAPR an intolerable characteristic. A modified OFDMA system, namely Fourier pre-coded OFDMA was proposed to solve the high PAPR problem in OFDMA. The Fourier pre-coded OFDMA (more commonly known as single carrier - frequency division multiple access (SC-FDMA)) retains the advantages of OFDMA, while eliminating the problem of high PAPR. Due to these characteristics, SC-FDMA has been adopted as the uplink multiple access scheme in 3GPP long term evolution (LTE) \cite{myung2006single}.

The wideband nature of SC-FDMA makes it highly susceptible to narrowband interference (NBI). The NBI sources include other devices operating in the same spectrum (e.g., cordless phones, garage openers etc.) and other communication systems operating in a cognitive manner. Here it is worth mentioning that though OFDMA is equally susceptible to these NBI sources, there is a fundamental difference in the way NBI affects the data in SC-FDMA and OFDMA. While a single NBI source (aligned with the grid of the system under consideration) affects only one sub-carrier in OFDMA, it perturbs all data points in SC-FDMA system. This makes NBI mitigation in SC-FDMA vital for reliable performance of the communication system. At high signal-to-interference ratio (SIR), coding can be relied on to mitigate the errors introduced by the NBI. However, at low SIR levels, interference begins to overwhelm the code and necessitates a receiver that is able to directly deal with it.

In this work, we exploit the sparse nature of the NBI to recover it using a low complexity Bayesian sparse reconstruction procedure. Specifically, we utilize the support agnostic Bayesian matching pursuit (SABMP) algorithm (proposed by some of the authors in \cite{masood2013sparse}) for NBI recovery. The SABMP algorithm uses the statistics of additive noise (which is assumed Gaussian), but is agnostic to the distribution of the active elements. This characteristic plays a vital role in NBI-impaired signal restoration as the distribution of the NBI signal might not be known. Further, the practical scenario of grid mismatch is also considered and the spreading effect is more realistically modelled by allowing the various NBI sources to have independent grid offsets. It is noted that the spectral spillover caused by the grid mismatch destroys the sparsity of the unknown signal. A well-accepted methodology to spectrally contain the spread NBI is windowing \cite{redfern2002receiver}. However, in this work, we use the \emph{Haar}  transform to sparsify the NBI, which is shown to outperform windowing in this aspect. Due to the devastating effect of the NBI in low SIR regime, we presume (throughout this work) that sparing a small subset of data points for sensing the NBI is a reasonable choice. Moreover, to minimize the number of reserved tones (and hence to maximize the spectral efficiency) a data-aided NBI mitigation technique is proposed. Using the proposed data-aided technique, the receiver probabilistically assigns a confidence level to each data point. A few data points (with highest confidence levels) are then selected and used in conjunction with reserved tones to enhance the NBI estimation accuracy. Finally, we extend the proposed reconstruction scheme to the multiple antenna context. This extension is motivated by the observation that NBI on each receive antenna will have the \emph{same} support and possibly different magnitudes and phases. Hence, the antennas can collaboratively estimate the support of NBI signal to improve the estimation accuracy.

The proposed scheme is distinguishable from existing literature as it aims at a general scenario of time-varying (changing completely from symbol-to-symbol) multiple NBI sources with independent grid offsets. Note that, several studies considered the impact of NBI in multi-carrier systems and numerous strategies have been devised. Available NBI mitigation schemes commonly adopt one of the following three methodologies: \emph{avoidance} \cite{zhang2010robust,wang2007spectrum,coon2008narrowband}, \emph{spreading} \cite{gerakoulis2002interference,wu2005narrowband} and \emph{subtraction} \cite{darsena2007successive,darsena2008successive,gomaa2011sparsity,sohail2012narrow}. However, these schemes are designed for OFDMA and do not readily apply to SC-FDMA as the two systems are fundamentally different. The literature addressing the problem of NBI specifically for SC-FDMA is seriously limited and only a handful of articles are available (e.g., \cite{mei2013wfrft,ccelebi2013interference}). Furthermore, these articles address specific cases (e.g., single NBI sources that don't change much over multiple symbols) under idealistic assumptions (e.g., known power and location). In this relation, the proposed scheme completely relaxes the requirement of known power (assumed in \cite{mei2013wfrft,sohail2012narrow}) and known location (assumed in \cite{mei2013wfrft,ccelebi2013interference}). Further, owing to multiple interferers, we consider that any sub-carrier within SC-FDMA band is susceptible to NBI, unlike \cite{ccelebi2013interference} that assumes consecutive impaired tones. We would also like to highlight difference between the proposed scheme and \cite{gomaa2011sparsity,sohail2012narrow} (i.e., existing works that exploit sparsity of NBI for its estimation in zero padded - OFDM). Gomma and Al-Dhahir \cite{gomaa2011sparsity}, opted for $\ell_1$-optimization based recovery of the unknown signal, which is very complex for real time implementation. To reduce the computational burden, Sohail \emph{et al.} \cite{sohail2012narrow} utilized the prior structural information and performed maximum a posteriori estimation assuming Gaussian prior on the unknown and availability of second order statistics. In contrast, we propose a low complexity Bayesian recovery scheme that is agnostic to the distribution of the unknown and does not require the statistics of the signal.

The main contributions of this work can be summarized as follows:
\begin{enumerate}
\item An NBI mitigation scheme is proposed that targets multiple time-variant NBI sources with independent grid offsets.
\item A low complexity, sparsity aware, Bayesian NBI reconstruction methodology is proposed. The proposed Bayesian method is agnostic to the distribution of the NBI.
\item A realistic model for grid mismatch is used that allows the NBI sources to have independent grid offsets.
\item Haar wavelet transformation is utilized to sparsify the unknown spread NBI signal.
\item A data-aided approach for NBI recovery is presented to improve the spectral efficiency of the proposed scheme.
\item The proposed scheme is extended to single-input multi-output (SIMO) systems by exploiting the joint-sparsity of NBI signals over all antenna elements.
\end{enumerate}
\subsection{Paper Organization}
The remainder of the paper is organized as follows. Section~\ref{sec:DatModProbForm} introduces the data model for NBI impaired SC-FDMA transmission. To mitigate the NBI, a Bayesian sparse recovery procedure is presented in Section~\ref{sec:BSRNBI}. The data-aided NBI recovery procedure is outlined in Section~\ref{sec:ANBIR}. Section~\ref{sec:MABS} extends the proposed NBI recovery scheme to SIMO systems and finally Section~\ref{sec:Conc} concludes the paper. To avoid a bulky simulation section at the end, each section is made self contained in terms of numerical results. We set the stage by introducing our notation.
\subsection{Notation}
Unless otherwise noted, scalars are represented by italic letters (e.g. $N$). Bold-face lower-case letters (e.g. $\xv$) are reserved to denote time domain vectors, and frequency domain vectors are represented using bold-face upper-case calligraphic letters (e.g. $\Xfd$). Bold-face upper-case letters are associated with matrices (e.g $\Xm$). The symbols $\hat\xv$, $\xv^\herm$, $x(i)$ and $x^\ast(i)$ represent the estimate, hermitian (conjugate transpose), $i{\rm th}$ entry and the conjugated $i{\rm th}$ entry of the vector $\xv$. The cardinality of a set $\Tc$ will be denoted by $|\Tc|$. Further, $\EE[\cdot]$, $\Id$ and $\zerov$ denote the expectation operator, identity matrix and the zero vector, respectively.
\section{SC-FDMA and NBI Model}\label{sec:DatModProbForm}
Consider an uplink SC-FDMA system with $U$ users. In such a system, the $u{\rm th}$ user converts the incoming high rate bit stream into $P$ parallel streams. These low rate bit streams are then modulated using a $Q$-ary QAM alphabet $\{\Ac_0,\Ac_1,\cdots,\Ac_{Q-1}\}$, resulting in a $P$ dimensional data vector $\xv_u$. The data $\xv_u$ is Fourier pre-coded using the $P\times P$ discrete Fourier matrix $\Fm_P$ to lower the PAPR of the transmission signal. The $(k,l){\rm th}$ element of $\Fm_P$ is given by
\begin{align}
\!\!\!f_P(k,l)\!=\!P^{-1/2}\exp\left(-\jmath \frac{2\pi kl}{P}\right),~k,l\in0,1,\cdots,P-1.
\end{align}
The pre-coded data $\Fm_P\xv_u$ is now mapped to the sub-carriers designated for $u{\rm th}$ user. The sub-carrier/resource allocation can be done in a localized or distributed manner (see \cite{myung2006single} for details). In this work, we only consider \emph{interleaved} SC-FDMA (i.e., SC-FDMA with interleaved resource allocation). The motivation behind the use of interleaved allocation is the robustness of this setting to frequency selective fading \cite{myung2006single}. For the $u{\rm th}$ user, the data $\Fm_P\xv_u$ is mapped to the designated sub-carriers by using an $N\!\times\!P$ ($N=PU$) resource allocation matrix $\Mm_u$. For interleaved assignment, the $(k,l){\rm th}$ element of $\Mm_u$ is given by
\begin{align}
m_u(k,l)=
\begin{cases}
1,&k=(u-1)+Ul,~~0\leq l \leq P-1,\\
0,& \text{otherwise}.
\end{cases}
\end{align}
This makes resource allocation matrices belonging to different users orthonormal, i.e.,
\begin{align}\label{eqsamtxproperty}
\Mm_{i}^\herm\Mm_{j}=
\begin{cases}
\Id_P,&i=j,\\
\zerov_P,&i\neq j.
\end{cases}
\end{align}
Now, the $N$ dimensional inverse DFT (IDFT) operation (i.e., $\Fm_N^\herm$) on $\Xfd_u=\Mm_u\Fm_P\xv_u$ results in the desired time domain transmission signal. After adding the cyclic prefix, the time domain signal is fed to a finite impulse response channel of length $N_c$, $\hv_u=[h_u^\ast(0),h_u^\ast(1),\cdots,h_u^\ast(N_c-1)]^\herm$. The channel tap coefficients form a zero mean, complex Gaussian, independent and identically distributed (i.i.d) collection. We assume perfect time and frequency synchronization between mobile terminals and the base-station (BS). Hence, after removing the cyclic prefixes, the received time domain signal (in absence of NBI) can be written as
\begin{align}\label{eqrecsigtime}
\yv=\sum_{u=0}^{U-1}\Hm_u\Fm_N^\herm\Xfd_u+\zv,
\end{align}
where $\Hm_u$ is the circulant channel matrix for $u{\rm th}$ user and $\zv$ is the additive white Gaussian noise (AWGN) with $\zv\sim\Cc\Nc(\zerov,\sigma_z^2\Id_N)$. The circulant nature of $\Hm_u$ allows us to diagonalize it using DFT matrix $\Fm_N$ and write $\Hm_u=\Fm_N^\herm\Lambdam_u\Fm_N$, where $\Lambdam_u$ is a diagonal matrix with channel frequency response on its diagonal. In this work, the channel impulse response is assumed known at the receiver and hence $\Hm_u$ and $\Lambdam_u$ are readily available. The frequency domain received data vector $\Yfd$ is now given by
\begin{align}\label{eqrecsigfreq}
\Yfd=\Fm_N\yv=\sum_{u=0}^{U-1}\Lambdam_u\Xfd_u+\Zfd,
\end{align}
where $\Lambdam_u=\Fm_N\Hm_u\Fm_N^\herm$ and $\Zfd=\Fm_N\zv$. Utilizing (\ref{eqsamtxproperty}) and the diagonal nature of $\Lambdam_u$, the data vector $\xv_u$ can be estimated by using the zero forcing - frequency domain equalization (ZF-FDE). The ZF-FDE is implemented by projecting $\Yfd$ on $\Fm_P^\herm\Mm_u^\herm\Lambdam_u^{-1}$ to get
\begin{align}\label{eqestnonbi}
\hat{\xv}_{u,{\rm ZF}}=\xv_u+\Fm_P^\herm\Mm_u^\herm\Lambdam_u^{-1}\Zfd.
\end{align}

Though ZF-FDE is a reasonable choice for milder channels, it is not suitable when the frequency response contains nulls. This is because the noise corresponding to a spectral null (i.e., a weak channel) is greatly enhanced upon applying the ZF-FDE. Further, as the enhanced noise (i.e., $\Lambdam_u^{-1}\Zfd$) impacts the data $\xv_u$ through the IDFT operation $\Fm_P^\herm$, a single spectral null can considerably increase the bit error rate (BER)\footnote{In Section~\ref{sec:NC}, the proposed sparse reconstruction scheme is explored from an alternative viewpoint of enhanced noise-cancellation in ZF-FDE regime.} \cite{huang2008decision}. To address this issue, minimum mean square error - FDE (MMSE-FDE), turbo equalizers and decision feedback equalizers are explored as replacements for ZF-FDE \cite{wang2008performance,berardinelli2008improving,huang2008decision}. In this work, we use MMSE-FDE to obtain the following estimate
\begin{align}\label{eqestnonbiMMSE1}
\hat{\xv}_{u,{\rm MMSE}}=\Rm_{\xv}\Am^\herm\left(\Am\Rm_{\xv}\Am^\herm+\sigma_z^2\Id\right)^{-1}\Mm_u^\herm\Yfd,
\end{align}
where $\Rm_{\xv}\triangleq\EE[\xv_u\xv_u^\herm]=\sigma_x^2\Id$ is the auto-correlation matrix of the data vector and $\Am\triangleq\Mm_u^\herm\Lambdam_u\Mm_u\Fm_P$. Both ZF and MMSE estimators are linear in $\Yfd$, hence we can simply write $\hat{\xv}_u=\Em_u\Yfd$ (dropping the subscripts ZF and MMSE, which can be understood from the context), where
\begin{align}
\Em_u=
\begin{cases}
\Fm_P^\herm\Mm_u^\herm\Lambdam_u^{-1}&\text{ZF}\\
\sigma_x^2\Am^\herm\left(\sigma_x^2\Am\Am^\herm+\sigma_z^2\Id\right)^{-1}\Mm_u^\herm&\text{MMSE}
\end{cases}
\end{align}

Using the definition of $\Em_u$, we can write $\hat{\xv}_u=\xv_u+\Em_u\Zfd$, which is true, exactly for ZF-FDE and approximately for MMSE-FDE as $\Em_u\Lambdam_u\Mm_u\Fm_P\approx\Id$ (the approximation tends to equality as $\sigma_z^2\rightarrow 0$).

Though (\ref{eqestnonbiMMSE1}) provides a good estimate of $\xv_u$ in the NBI free regime, it is not suitable for systems experiencing NBI. In the following subsection, we explain how NBI affects the SC-FDMA system.
\subsection{The NBI Impaired SC-FDMA}
The received SC-FDMA signal might be impaired by a single or multiple time-variant NBI sources. Let $\Ifd_L$ be an $L$ dimensional vector representing the active NBI sources. Using $\Ifd_L$, we obtain an $N$ dimensional NBI signal $\Ifd=\Fm_N\bar{\Fm}_N^\herm\Ifd_L$, where $\bar{\Fm}_N^\herm$ is an $N \times L$ partial IDFT matrix containing the columns corresponding to the frequencies of active NBI sources. Here, it is important to understand that channels between the NBI sources and the BS are absorbed into $\Ifd_L$. In other words, we can say that $\Ifd_L=\Lambdam_{\boldsymbol{\bar{\Ic}}_L}\bar\Ifd_L$, where $\Lambdam_{\boldsymbol{\bar{\Ic}}_L}$ is a diagonal $L\!\times\!L$ matrix containing the frequency domain channel gains between the interference sources and the receiver antennas and $\bar\Ifd_L$ represents the actual interference sources\footnote{The sparsity of unknown NBI source can be preserved in effective NBI if the frequency domain NBI channel matrix is diagonal. In this relation, our work parallels prior works on sparse NBI mitigation that considered the time domain channel between the NBI source and the receiver to be positive semi-definite Hermitian Toeplitz and approximated it as circulant. This implies diagonal nature of NBI channel response matrix (see \cite{gomaa2011sparsity,sohail2012narrow} and references therein) and allows to estimate effective NBI that matches the sparsity of NBI source.}. Hence, a simple addition of $\Ifd$ in (\ref{eqrecsigfreq}) will yield the NBI impaired SC-FDMA received signal. This received signal is given as
\begin{align}\label{eqrecNBIsignal}
\Yfd=\sum_{u=0}^{U-1}\Lambdam_u\Xfd_u+\Ifd+\Zfd.
\end{align}

In practice, the NBI sources may have a grid offset with the SC-FDMA system, causing energy of the NBI to spill over all tones. A spreading matrix $\Hm_{fo}=\Fm_N\Lambdam_{fo}\Fm_N^\herm$ is commonly used to model grid offset between the NBI signal and the system under consideration \cite{gomaa2011sparsity,sohail2012narrow}. The diagonal matrix $\Lambdam_{fo}$ is defined as $\Lambdam_{fo}\triangleq\diag(1,\exp(\jmath \frac{2\pi\alpha(1)}{N},\cdots,\exp(\jmath \frac{2\pi\alpha(N-1)}{N}))$, where $\alpha$ is a random number uniformly distributed over the interval $[-\frac{1}{2},\frac{1}{2}]$. A fundamental limitation of this model is its inability to assume independent grid offsets for multiple NBI sources. To overcome this limitation, we define the spread NBI signal as
\begin{align}\label{eqNBImodel}
\Ifd=\Fm_N\bar{\Fm}_{con}^\herm\Ifd_L,
\end{align}
where $\bar{\Fm}_{con}$ is the $L\!\times\!N$ continuous DFT matrix, with $(f_l,k){\rm th}$ entry
\begin{align}\label{eqgridoffsetfreq}
\!\bar{\Fm}_{con,(f_l,k)}\!=\!N^{-1/2}\exp\left(-\jmath \frac{2\pi f_l k}{N}\right),
\begin{aligned}
l&\in 0,1,\cdots,L-1,\\
k&\in 0,1,\cdots,N-1.
\end{aligned}
\end{align}
As the normalized frequencies $f_l/N\in[0,1)$ are drawn independently, they emulate independent grid offsets for different NBI sources. Recently, Tang \emph{et al.} used a similar modelling approach in an attempt to estimate continuous frequencies and amplitudes of a mixture of complex sinusoids \cite{tang2013compressed}.

The estimate of the transmitted signal $\xv_u$ in NBI free case (i.e., (\ref{eqrecsigfreq})) is obtained using (\ref{eqestnonbiMMSE1}). However, following the same estimation procedure for NBI impaired system (i.e., (\ref{eqrecNBIsignal})) will yield
\begin{align}\label{eqestnbi}
\hat{\xv}_u=\xv_u+\Em_u(\Ifd+\Zfd),
\end{align}
which is not a reliable estimate of $\xv_u$ due to the presence of $\Ifd$. Further, note that $\Ifd$ perturbs $\xv_u$ through an IDFT operation (as evident by giving a closer look to the construction of $\Em_u$), hence, even in the optimistic case (i.e., a single NBI source with no grid offset) all data points are corrupted by the NBI. In low SIR scenarios, the interference might be strong enough to take a majority of data symbols out of their correct decision regions, resulting in an intolerably high BER. Thus, our task is the estimation/mitigation of $\Ifd$, which we pursue using a Bayesian sparse recovery framework.
\section{Bayesian Sparse Recovery of the NBI}\label{sec:BSRNBI}
To reconstruct the unknown NBI signal, we keep a randomly chosen subset of the vector $\Xfd_u$ data free and index this subset using $\Tc_u$. To extract the portion of the received signal corresponding to the reserved tones, let us define a $|\Tc_u|\!\times\!P$ binary selection matrix $\Sm_{\Tc_u}$. The selection matrix $\Sm_{\Tc_u}$ has one entry equal to $1$ per row, corresponding to the location of a reserved data point (with all other entries being zero). Now we proceed by projecting $\hat{\xv}_u$ (defined in (\ref{eqestnbi})) onto the subspace spanned by the reserved points, i.e.,
\begin{align}\label{equndetsys}
\underbrace{\Sm_{\Tc_u}\hat{\xv}_u}_{\xv_{u,\Tc}^\prime}&=\Sm_{\Tc_u}\xv_u+\underbrace{\Sm_{\Tc_u}\Em_u}_{\Psim_{u,\Tc}}\underbrace{(\Ifd+\Zfd)}_{\Ifd^\prime},\nonumber\\
\Longrightarrow\xv_{u,\Tc}^\prime&=\Psim_{u,\Tc}\Ifd^\prime,
\end{align}
where $\Sm_{\Tc_u}\xv_u=\zerov$. Owing to the presence of $\Em_u$ in the sensing matrix $\Psim_{u,\Tc}$, the columns corresponding to subcarriers assigned to user $u$ are the only nonzero columns of $\Psim_{u,\Tc}$. This fact has two important implications:
 \begin{enumerate}
 \item Only the portion of NBI falling on subcarriers allocated to user $u$ is projected on the measurement vector $\xv_{u,\Tc}^\prime$ and hence the subsystems (\ref{equndetsys}) corresponding to different users are uncoupled in terms of the information that they contain regarding the NBI. Further, the dimensionality of unknown can be reduced by eliminating zero columns of the sensing matrix.
 \item It does not serve any purpose to estimate NBI for all users jointly as i) subsystems (\ref{equndetsys}) belonging to each user are uncoupled, hence the joint estimate is unlikely to be more informative than individual estimate and ii) BS could be interested in NBI recovery for only a few users.
 \end{enumerate}

At this stage, we drop the subscript $u$ for notational convenience and simply write\footnote{Here onwards, the subscript $u$ is added (resp. removed) as per requirement (resp. notational convenience).} $\xv_\Tc^\prime=\Psim_\Tc\Ifd^\prime$. To recover $\Ifd$, the above under-determined system of equations can be solved using any compressed sensing (CS) reconstruction algorithm (e.g., \cite{candes2008introduction,ji2008bayesian,tropp2007signal,blumensath2008gradient,needell2009cosamp}). In this work, we follow a Bayesian sparse recovery framework for the estimation of the unknown NBI signal. However, a couple of fundamental challenges surface when talking about Bayesian sparse NBI recovery. The first challenge appears as common Bayesian approaches assume a known prior on the active elements of the unknown signal (see e.g., \cite{ji2008bayesian,babacan2010bayesian}), and we may not know the distribution of the NBI. The second challenge is the spreading of the NBI signal (due to grid offset) that destroys the sparsity of the unknown signal. These problems are addressed below.

\subsection{Prior on $\Ifd^\prime$}
It is a common practice in Bayesian schemes to assume a known prior on the unknown signal, e.g., \cite{babacan2010bayesian} assumes a Laplacian prior. However, recently Masood and Al-Naffouri proposed SABMP, a Bayesian scheme that is agnostic to the distribution of active taps, but acknowledges the sparsity of the unknown vector and Gaussianity of the additive noise \cite{masood2013sparse}. Further, the proposed scheme has been shown to outperform many algorithms, both for reconstruction accuracy and computational complexity (see \cite{masood2013sparse} for details). A brief description of SABMP algorithm is given in Appendix~\ref{AppA}. The agnostic nature of SABMP plays a vital role in NBI recovery as i) we may not know the distribution of $\Ifd$ and ii) even if we did know the distribution, it might be difficult to estimate its parameters (i.e., moments). Towards this end, let us recall that $\Ifd_L$ represents the joint channel-NBI source i.e., $\Ifd_L=\Lambdam_{\boldsymbol{\bar{\Ic}}_L}\bar\Ifd_L$. Here, an appropriate treatment would be to assume circularly symmetric complex Gaussian prior for both $\bar\Ifd_L$ and $\Lambdam_{\boldsymbol{\bar{\Ic}}_L}$. This implies that the entries of $\Ifd_L$ are formed by the product of two independent complex normal random variables. O'Donoughue and Moura coined the term \emph{complex Double Gaussian} for such a distribution \cite{o2012product}. Hence, in this case, though the distribution is known, its parameter estimation is relatively difficult. Further, if non-Gaussianity is assumed on the NBI-BS channel model, it may yield more complex statistical behaviour for $\Ifd_L$. As we are interested in recovering $\Ifd$, we note that for no grid offset, the active elements of $\Ifd$ will assume the distribution of $\Ifd_L$. However, grid offset will make the statistical characterization of $\Ifd$ even more challenging. For these reasons, a suitable reconstruction scheme would be able to work regardless of the distribution of unknown signal and whether this distribution is known or not. As the SABMP algorithm possesses these qualities and incurs low computational complexity, we employ SABMP as a sparse reconstruction scheme for NBI mitigation. In addition, the extension of the SABMP algorithm for multiple measurement vectors (MMV), namely MMV-SABMP \cite{al2013distribution}, is tailor-made to exploit the joint-sparsity of NBI signal in SIMO systems (see Section~\ref{sec:MABS} further ahead, which shows how to utilize multiple antennas at the BS for enhanced NBI estimation).
\subsection{Sparsifying $\Ifd^\prime$}
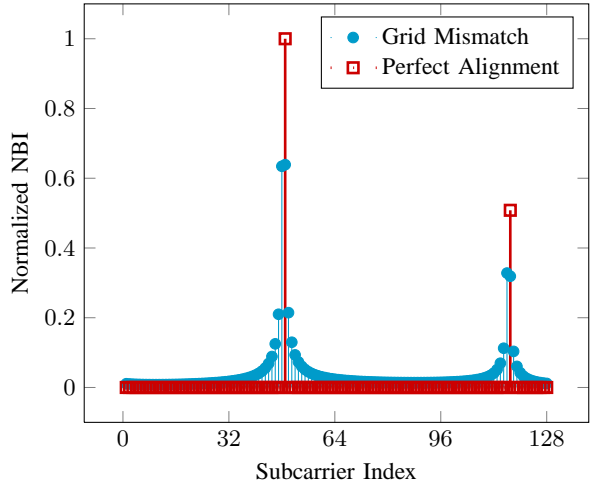
\begin{figure}[h!]\centering
{\small\begin{tikzpicture}
\begin{axis}[
xlabel={Subcarrier Index},
ylabel={Normalized NBI},
ylabel near ticks,
yminorticks,
xlabel near ticks,
xtick={0,32,64,96,128},
ytick={},
width=0.45\textwidth,
legend cell align=left,
legend pos=north east,
legend columns=1,
legend entries={Grid Mismatch, Perfect Alignment},
]
\addplot[mark=*,black!20!cyan,ycomb] table [x=Number, y=Off, col 
sep=comma] {spread.dat};
\addplot[mark=square,black!20!red,ycomb,line width=1pt] table [x=Number, y=NoOff, col 
sep=comma] {spread.dat};
\end{axis}
\end{tikzpicture}%
}
\caption{NBI spreading for two active NBI sources as a result of grid mismatch between the NBI sources and the SC-FDMA system.}
\label{fig:spread}
\end{figure}

A fundamental requirement of sub-Nyquist sampling based reconstruction (as pursued in this work) is the sparsity of the unknown signal. Though there are only a few active NBI sources, in practice, the non-orthogonality of these sources to the SC-FDMA grid destroys the frequency domain sparsity of the unknown signal (see Fig.~\ref{fig:spread}). In this subsection, we discuss how the sparsity of the spread NBI signal can be restored.

Two strategies are followed in literature to tackle the grid offset problem. One possibility is to estimate the gird offset (see e.g., \cite{zhu2011sparsity,zhu2011weighted}). The problem with offset estimation is that offset is a highly nonlinear function of the observations $\Yfd$. Further, the grid offset estimation is complicated by the fact that different NBI sources assume independent grid offsets. The second approach is more mainstream and directly deals with an NBI signal experiencing energy spill-over (due to the grid mismatch) by \emph{windowing} \cite{gomaa2011sparsity}. A windowing matrix function $\Hm_{win}=\Fm_N\Lambdam_w\Fm_N^\herm$ applied to the received signal sparsifies the unknown vector $\Ifd^\prime$. Here, $\Lambdam_w\!\triangleq\!\diag(w(0),w(1),\cdots,w(N-1))$ and $w(n)$ is the $n{\rm th}$ sample of the window function. It is a common practice to window the received time domain signal before taking the DFT. However, since the sole purpose of introducing windowing is enhancing the sparsity of $\Ifd^\prime$, we can postpone its inclusion till NBI reconstruction. To incorporate the windowing matrix function at NBI recovery stage we can re-write (\ref{equndetsys}) as
\begin{align}\label{eqwinform}
\xv_\Tc^\prime=\Psim_\Tc \Hm_{win}^{-1}\Hm_{win}\Ifd^\prime,
\end{align}
where we assume the non-singularity of $\Hm_{win}$. Now, if we sense via $\Psim_\Tc \Hm_{win}^{-1}$, we will be reconstructing $\Hm_{win}\Ifd^\prime$, which is more sparse compared to $\Ifd^\prime$. As the formulation (\ref{eqwinform}) requires only the non-singularity of $\Hm_{win}$, we are motivated to look for other possibilities towards sparsifying $\Ifd^\prime$. Our drive to seek a better replacement for $\Hm_{win}$ also stems from the fact that $\Hm_{win}$ is not a unitary matrix and hence lacks a very desirable property pertaining to dictionary design in standard CS \cite{davenport2013signal}. Speaking in terms of time and frequency domains, as the signal $\Ifd^\prime$ is no longer sparse in either, we seek another domain that has a sparse representation of $\Ifd^\prime$. Any transformation matrix that is; i) linear, ii) non-singular, iii) unitary and iv) a good choice for sparsifying NBI, will serve the purpose. While choosing a sparsifying transform for NBI reconstruction, though properties i), ii) and iii) will be promptly evident, property iv) needs some consideration. To this end, note that unlike sparse signals, \emph{compressible signals} (such as the NBI under grid offset) cannot be compared using $\ell_0$ norm. As $\|\Ifd^\prime\|_{\ell_0}=\|\Hm_{win}\Ifd^\prime\|_{\ell_0}=N$, counting the number of active elements will yield a false conclusion that windowing did not enhance the sparsity of the unknown. As practical signals are seldom sparse, sparsity measures other than $\|\!\cdot\!\|_{\ell_0}$ e.g., \emph{Gini index} (GI) \cite{zonoobi2011gini} and \emph{numerical sparsity} \cite{lopes2013estimating} have been put forth to compare compressible signals. In this work, we use GI (a normalized measure of sparsity) to compare sparsifying transforms. Consider a vector $\underline{\Ifd}^\prime=[\underline{\Ic}^\prime(0),\underline{\Ic}^\prime(1),\cdots,\underline{\Ic}^\prime(N-1)],$ with its elements re-ordered, such that $|\underline{\Ic}^\prime(0)|<|\underline{\Ic}^\prime(1)|<,\cdots,<|\underline{\Ic}^\prime(N-1)|$. The GI is then defined as
\begin{align}\label{GI}
\text{GI}(\underline{\Ifd}^\prime)=1-2\sum_{k=0}^{N-1}\dfrac{|\underline{\Ic}^\prime(k)|}{\|\underline{\Ifd}^\prime\|_{\ell_1}}
\left(\dfrac{N-k-\frac{1}{2}}{N}\right),
\end{align}
where $\|\!\cdot\!\|_{\ell_1}$ represents the $\ell_1$ norm. An important advantage of GI over the conventional norm measures is that it is normalized, and assumes values between $0$ and $1$ for any vector. Further, it is $0$ for the least sparse signal with all the coefficients having an equal amount of energy and $1$ for the most sparse signal which has all the energy concentrated in just one coefficient (see \cite{zonoobi2011gini} for details). Our numerical findings based on GI suggest that (among the tested transforms) \emph{Haar wavelet} transform \cite{haar1910theorie} maximizes the GI and also satisfies the other three required properties (i.e., it is linear, non-singular and unitary). As the discussion on all the tested transforms will take us too far afield, we will confine our attention to the sparsifying ability of the Haar transform in comparison with windowing. The unitary Haar transform $\Hm_{haar}$ can be applied to $\Ifd^\prime$ in a manner identical to (\ref{eqwinform}), i.e.,
\begin{align}\label{eqhaarform}
\xv_\Tc^\prime=\Psim_\Tc \Hm_{haar}^\herm\Hm_{haar}\Ifd^\prime,
\end{align}
where $\Hm_{haar}^\herm=\Hm_{haar}^{-1}$. Note that, in (\ref{eqhaarform}) the sensing matrix $\Psim_\Tc \Hm_{haar}^\herm$ does not contain zero columns, hence, unlike the perfect grid alignment case, dimensionality of the unknown cannot be reduced.
\subsection{Noise-cancellation in ZF-FDE Systems}\label{sec:NC}
So far the focus of our discussion has been the sparse NBI recovery in SC-FDMA systems under MMSE-FDE setup. However, in this subsection we take the liberty of a slight diversion from the main route and explore the proposed sparse recovery procedure as a noise-cancellation scheme in ZF-FDE. The ZF-FDE regime suffers from noise enhancement owing to the weak channels and unlike OFDMA (where a spectral null destroys the data on only one sub-carrier), the enhanced noise in SC-FDMA influences all data points. Hence, even a single spectral null can considerably elevate the BER. The sparse reconstruction procedure explained earlier can be put in use to solve this noise enhancement problem. Let us recall from (\ref{eqestnonbi}) that with ZF-FDE, we have $\hat{\xv}_u=\xv_u+\Fm_P^\herm\Mm_u^\herm\Lambdam_u^{-1}\Zfd$. Following the steps that led to (\ref{equndetsys}), we continue by projecting (\ref{eqestnonbi}) on $\Sm_\Tc$ and establish some notational simplifications to get
\begin{align}\label{eqnoiseprob}
\xv^\prime_{\Tc}=\Psim_{\Tc}\Zfd
\end{align}
where $\Sm_\Tc\xv_u=\zerov$, $\xv^\prime_{\Tc}=\Sm_\Tc\hat{\xv}_{u}$ and $\Psim_{\Tc}=\Sm_\Tc\Fm_P^\herm\Mm_u^\herm\Lambdam_u^{-1}$. Now, we intend to solve (\ref{eqnoiseprob}) for the unknown $\Zfd$ using SABMP. However, by the very nature of sparse signal reconstruction, we are limited to reconstruct only a part of $\Zfd$. Luckily, in solving (\ref{eqnoiseprob}) using SABMP, the portion of $\Zfd$ that is reconstructed falls exactly on the weak channels (i.e., on the locations where the impact of noise is most pronounced). The sparse noise reconstruction on weak channels is possible as the sensing matrix $\Psim_{\Tc}$ contains the inverse channel $\Lambdam^{-1}$. To comprehend this assertion, note that the presence of $\Lambdam^{-1}$ significantly enhances the $\Psim_{\Tc}$ columns corresponding to the spectral nulls. As the measurement vector $\xv_{\Tc}^\prime$ is a linear combination of the columns of $\Psim_{\Tc}$, the stronger columns have a significant contribution in $\xv_{\Tc}^\prime$. In other words, the entries of $\Zfd$ corresponding to the weak channels are seen much more prominently in $\xv_{\Tc}^\prime$, and hence have a high probability of recovery. Recall that it is the noise at exactly these weak channels that inflates the BER. Now, that this noise is recovered and taken out, the noise enhancement problem is resolved to a great extent.
\subsection{Simulation Results}\label{ref:simres1}
A $512$ sub-carrier SC-FDMA system is simulated, with $2$ active users accessing the frequency resources in interleaved manner. The channel delay spread is quarter the symbol duration i.e., $N_c=N/4$ and $16$-QAM modulation is utilized. The NBI vector $\Ifd_L$ is obtained from complex normal distribution with SIR=$-10$dB. Two experiments are conducted in order to demonstrate the ability of the proposed reconstruction scheme to successfully recover the NBI. In the first experiment no grid offset is assumed, whereas, the second experiment assumes the realistic grid mismatch case. The third experiment is conducted to examine the noise-cancellation ability of the proposed scheme for ZF-FDE setup. The performance of the proposed scheme is compared with $\ell_1$-optimization based NBI recovery \cite{gomaa2011sparsity}.
\subsubsection{Experiment 1: Reconstruction with no Grid Offset}
In this experiment, we let the number of active NBI sources vary from symbol to symbol with a maximum of four active NBI sources per symbol. The locations of the active NBI sources also vary, however, all NBI sources are restricted to fall on the grid. Fig.~\ref{fig:part1_exp1_1} presents the BER results as a function of energy per bit ($E_b/N_0$) with $64$ reserved tones per user (this corresponds to a sub-sampling rate $\frac{|\Tc|}{N}=\frac{64}{256}=\frac{1}{4}$).\footnote{The sub-sampling rate $1/4$ implies that $25\%$ of the data-carriers are used as measurement tones and hence this sub-sampling rate is referred to as $|\Tc|\%=25\%$.} The results depict the ability of the proposed scheme to effectively recover the NBI. Further, observe that the (\emph{proposed}) scheme is slightly inferior to the (\emph{NBI free}) case for lower $E_b/N_0$ as the approximation $\Em_u\Lambdam_u\Mm_u\Fm_P\approx\Id$ is weak. However, as $E_b/N_0$ increases, the approximation tends towards equality, and the performance of the (\emph{proposed}) scheme improves. Further, though the noise enhancement issue is taken care of (to a great extent) by using the MMSE-FDE, part of noise on relatively weaker channels is still reconstructed and hence the (\emph{proposed}) scheme performs better than (\emph{NBI free}) case for higher $E_b/N_0$. Note also that there is no visual difference between the reconstruction accuracy of the (\emph{proposed}) scheme and (\emph{$\ell_1$-optimization}). However, the subgraph depicting the average run-time of the (\emph{proposed}) scheme shows that (\emph{$\ell_1$-optimization}) incurs high implementation complexity. These findings are inline with the fact that the computational complexity incurred by the (\emph{proposed}) scheme is of the order $\Oc(|\Tc|sN^2)$ in comparison with $\Oc(N^3)$ for (\emph{$\ell_1$-optimization}), when an $N$ dimensional unknown with $sN$ nonzero elements is reconstructed using $|\Tc|$ measurements \cite{masood2013sparse}. Fig.~\ref{fig:part1_exp1_2} presents the BER performance as a function of reserved tones with $E_b/N_0$ fixed at $17.5$dB. These results depict that acceptable BER performance might not be achieved by choosing an arbitrarily small number of reserved tones. Further, the (\emph{$\ell_1$-optimization}) yields better performance (when $|\Tc|\%$ is low) at the expense of higher computational complexity.

At this stage, it is worth highlighting that the SABMP requires the estimate of the probability of active elements $\hat{s}$, whereas the $\ell_1$-optimization does not. Hence, we conduct an experiment to test the robustness of the SABMP algorithm to errors in the estimate of the sparsity. The results shown in Fig.~\ref{fig:part1_exp1_3} depict the BER performance of the proposed scheme as a function of the initial estimate of the sparsity rate $\hat{s}$. The estimate values are varied from $20\%$ of the true value (i.e., $0.2s$) to $180\%$ of the true value (i.e., $1.8s$). It is observed, that the (\emph{proposed}) scheme is relatively insensitive to the errors in the estimate of the sparsity rate, and the BER performance varies only slightly over the range of interest.
\begin{figure}[h!]\centering
{\small\begin{center}
\begin{tikzpicture}[baseline]
\begin{semilogyaxis}[
xlabel={$E_b/N_0$ (dB)},
ylabel={BER},
ylabel near ticks,
yminorticks,
xlabel near ticks,
xtick={5,10,15,20,25},
width=0.45\textwidth,
legend cell align=left,
legend to name=named11,
legend columns=-1,
legend entries={NBI Impaired,NBI Free, Proposed, $\ell_1$-optimization},
grid=major,
]
\addplot[mark=square,black!20!brown,line width=1pt] table [x=EbNo, y=NBI, col 
sep=comma] {Part1_Exp1_1.dat};
\addplot[mark=o,black!20!red,line width=1pt] table [x=EbNo, y=NoNBI, col 
sep=comma] {Part1_Exp1_1.dat};
\addplot[mark=diamond,black!20!blue,mark options={scale=1.3},line width=1pt] table [x=EbNo, y=SABMP, col 
sep=comma] {Part1_Exp1_1.dat};
\addplot[mark=triangle,black!40!green,mark options={scale=1.3},line width=1pt] table [x=EbNo, y=l1, col 
sep=comma] {Part1_Exp1_1.dat};
\end{semilogyaxis}
\begin{semilogyaxis}[xshift=15pt,yshift=15pt,
ylabel={Run-time},
yticklabel style={
rotate=90,},
ylabel absolute, 
ylabel style={yshift=-130pt},
yminorticks,
xtick={5,10,15,20,25},
width=0.25\textwidth,
]
\addplot[mark=diamond,black!20!blue,mark options={scale=1.3},line width=1pt] table [x=EbNo, y=SABMP, col 
sep=comma] {Part1_Exp1_1_RT.dat};
\addplot[mark=triangle,black!40!green,mark options={scale=1.3},line width=1pt] table [x=EbNo, y=l1, col 
sep=comma] {Part1_Exp1_1_RT.dat};
\end{semilogyaxis}
\end{tikzpicture}
\\
\ref{named11}
\end{center}}
\caption{BER performance of the proposed sparse NBI reconstruction scheme as a function of $E_b/N_0$ with $|\Tc|\%=25\%$ for perfect grid alignment}
\label{fig:part1_exp1_1}
\end{figure}
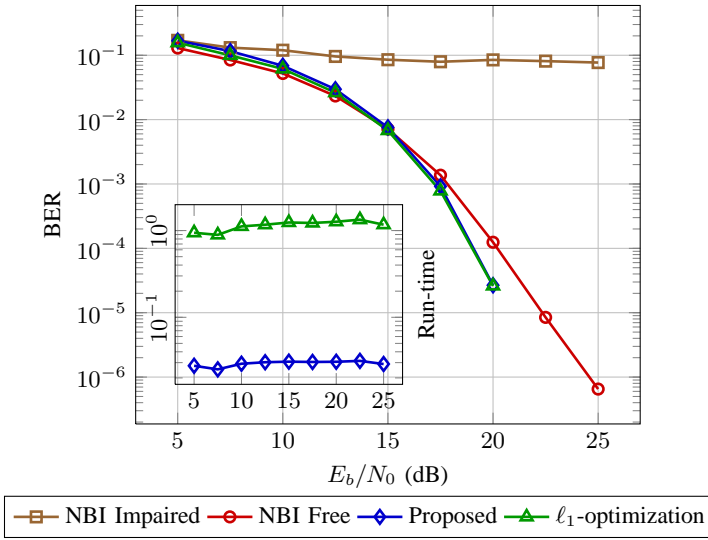
\begin{figure}[h!]\centering
{\small\begin{center}
\begin{tikzpicture}
\begin{semilogyaxis}[
xlabel={$|\Tc|\%$},
ylabel={BER},
ylabel near ticks,
yminorticks,
xlabel near ticks,
width=0.45\textwidth,
legend cell align=left,
legend to name=named12,
legend columns=-1,
legend entries={NBI Impaired,NBI Free, Proposed, $\ell_1$-optimization},
grid=major,
]
\addplot[mark=square,black!20!brown,line width=1pt] table [x=TS, y=NBI, col 
sep=comma] {Part1_Exp1_2.dat};
\addplot[mark=o,black!20!red,line width=1pt] table [x=TS, y=NoNBI, col 
sep=comma] {Part1_Exp1_2.dat};
\addplot[mark=diamond,black!20!blue,mark options={scale=1.3},line width=1pt] table [x=TS, y=SABMP, col 
sep=comma] {Part1_Exp1_2.dat};
\addplot[mark=triangle,black!40!green,mark options={scale=1.3},line width=1pt] table [x=TS, y=l1, col 
sep=comma] {Part1_Exp1_2.dat};
\end{semilogyaxis}
\begin{semilogyaxis}[xshift=85pt,yshift=70pt,
ylabel={Run-time},
yticklabel style={
rotate=90,},
ylabel absolute, 
ylabel style={yshift=-130pt},
yminorticks,
width=0.25\textwidth,
]
\addplot[mark=diamond,black!20!blue,mark options={scale=1.3},line width=1pt] table [x=TS, y=SABMP, col 
sep=comma] {Part1_Exp1_2_RT.dat};
\addplot[mark=triangle,black!40!green,mark options={scale=1.3},line width=1pt] table [x=TS, y=l1, col 
sep=comma] {Part1_Exp1_2_RT.dat};
\end{semilogyaxis}
\end{tikzpicture}
\\
\ref{named12}
\end{center}}
\caption{BER performance of the proposed sparse NBI reconstruction scheme as a function of $|\Tc|\%$ with $E_b/N_0=17.5$dB for perfect grid alignment.}
\label{fig:part1_exp1_2}
\end{figure}
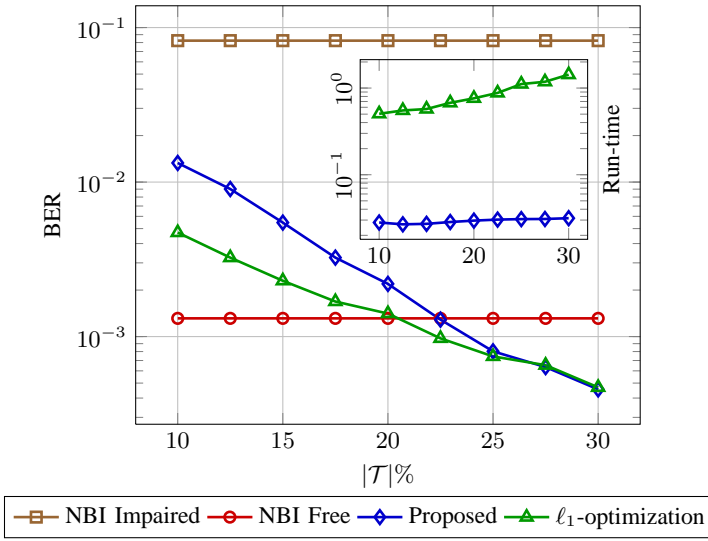
\begin{figure}[h!]\centering
{\small\begin{tikzpicture}
\begin{semilogyaxis}[
xlabel={$\hat{s}$},
ylabel={BER},
ylabel near ticks,
yminorticks,
xlabel near ticks,
xtick={0.2,0.6,1,1.4,1.8},
xticklabels={$0.2s$,$0.6s$,$1s$,$1.4s$,$1.8s$},
width=0.45\textwidth,
legend cell align=left,
legend style={at={(0.96,0.88)}},
legend columns=1,
legend entries={NBI Impaired, NBI Free, Proposed},
grid=major,
]
\addplot[mark=square,black!20!brown,line width=1pt] table [x=pvec, y=NBI, col 
sep=comma] {Part1_Exp1_3.dat};
\addplot[mark=o,black!20!red,line width=1pt] table [x=pvec, y=NoNBI, col 
sep=comma] {Part1_Exp1_3.dat};
\addplot[mark=diamond,black!20!blue,mark options={scale=1.3},line width=1pt] table [x=pvec, y=SABMP, col 
sep=comma] {Part1_Exp1_3.dat};
\end{semilogyaxis}
\end{tikzpicture}%
}
\caption{BER performance of the proposed sparse NBI reconstruction scheme as a function of the initial estimate of the sparsity rate $\hat{s}$ for perfect grid alignment.}
\label{fig:part1_exp1_3}
\end{figure}
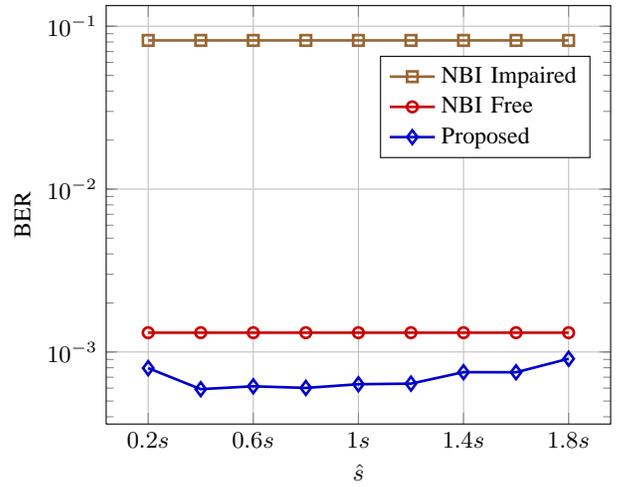
\subsubsection{Experiment 2: Sparsification using Haar Transform and Reconstruction Accuracy}
In this experiment, first we compare the Haar transform and windowing (Hamming \cite{gomaa2011sparsity}) for their sparsifying ability. The NBI sources are generated with
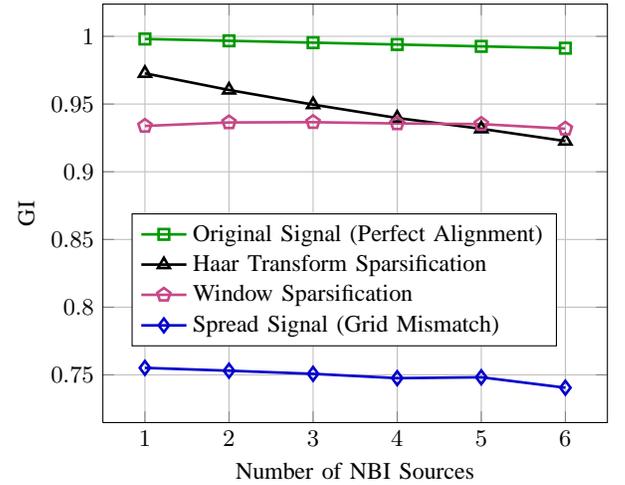
\begin{figure}[h!]\centering
{\small\begin{tikzpicture}
\begin{axis}[
xlabel={Number of NBI Sources},
ylabel={GI},
ylabel near ticks,
yminorticks,
xlabel near ticks,
ytick={0.75,0.8,0.85,0.9,0.95,1},
width=0.45\textwidth,
legend cell align=left,
legend style={at={(0.9,0.5)}},
legend columns=1,
legend entries={Original Signal (Perfect Alignment),Haar Transform Sparsification,Window Sparsification,Spread Signal (Grid Mismatch)},
grid=major,
]
\addplot[mark=square,black!40!green,line width=1pt] table [x=NumNBI, y=NoGrid, col 
sep=comma] {Part1_Exp2_1.dat};
\addplot[mark=triangle,black,mark options={scale=1.3},line width=1pt] table [x=NumNBI, y=Haar, col 
sep=comma] {Part1_Exp2_1.dat};
\addplot[mark=pentagon,black!20!magenta,mark options={scale=1.3},line width=1pt] table [x=NumNBI, y=Window, col 
sep=comma] {Part1_Exp2_1.dat};
\addplot[mark=diamond,black!20!blue,mark options={scale=1.3},line width=1pt] table [x=NumNBI, y=Spread, col 
sep=comma] {Part1_Exp2_1.dat};
\end{axis}
\end{tikzpicture}%
}
\caption{GI comparison (as a function of active NBI sources) of Haar transform and windowing based sparsity restoration averaged over $1000$ independent runs.}
\label{fig:part1_exp2_1}
\end{figure}
independent grid offset according to (\ref{eqNBImodel}). The GI is calculated (as a function of active NBI sources) and averaged over $1000$ independent runs for $\Ifd^\prime$ and its two transformed counterparts  ($\Hm_{win}\Ifd^\prime$ and $\Hm_{haar}\Ifd^\prime$). From the results (in Fig.~\ref{fig:part1_exp2_1}), we observe that for a small number of active NBI sources (i.e., $\leq 4$) the Haar transform has better sparsifying ability than windowing. Further, the BER performance of proposed SABMP reconstruction scheme for the cases of spread signal (\emph{spread}), windowing (\emph{window}) and Haar transform (\emph{Haar}) is shown in Fig.~\ref{fig:part1_exp2_2} and is compared with (\emph{$\ell_1$-optimization}) (as a function of $E_b/N_0$). The $\ell_1$-optimization based NBI recovery is performed using windowing sparsity restoration \cite{gomaa2011sparsity}. A maximum number of four active NBI sources with varying locations and independent frequency offsets per symbol are assumed with sub-sampling rate $1/4$. The lower BER for (\emph{Haar}) transform (in Fig.~\ref{fig:part1_exp2_2}) supports the conclusion that Haar transform possesses better sparsifying characteristics. The low BER is expected as in sparse reconstruction, a scheme better able to sparsify $\Ifd^\prime$ will yield better reconstruction accuracy and hence a lower BER. Further, it is noticed that with grid offset, windowed (\emph{$\ell_1$-optimization}) has an inferior performance to the proposed NBI reconstruction scheme. This behavior is expected, as the performance of $\ell_1$-optimization deteriorates with an increase in the sparsity rate. Fig.~\ref{fig:part1_exp2_3} shows the BER performance as a function of reserved tones. It is evident from the results that throughout the range of interest, Haar transform yields lower error rate as compared with (\emph{spread}) and (\emph{window}). Further, note that in addition to its superior BER performance, SABMP is able to run at a much lower complexity as compared to (\emph{$\ell_1$-optimization}) (as demonstrated in Fig.~\ref{fig:part1_exp1_2}).
\begin{figure}[h!]\centering
{\small\begin{center}
\begin{tikzpicture}[baseline]
\begin{semilogyaxis}[
xlabel={$E_b/N_0$ (dB)},
ylabel={BER},
ylabel near ticks,
yminorticks,
xlabel near ticks,
xtick={5,10,15,20,25},
width=0.45\textwidth,
legend cell align=left,
legend pos=south west,
legend columns=1,
legend entries={NBI Impaired, $\ell_1$-optimization, Spread, Window, Haar, NBI free},
grid=major,
]
\addplot[mark=square,black!20!brown,line width=1pt] table [x=EbNo, y=NBI, col 
sep=comma] {Part1_Exp2_2.dat};
\addplot[mark=x,black!40!green,mark options={scale=1.3},line width=1pt] table [x=EbNo, y=l1, col 
sep=comma] {Part1_Exp2_2.dat};
\addplot[mark=diamond,black!20!blue,mark options={scale=1.3},line width=1pt] table [x=EbNo, y=SABMP, col 
sep=comma] {Part1_Exp2_2.dat};
\addplot[mark=pentagon,black!20!magenta,mark options={scale=1.3},line width=1pt] table [x=EbNo, y=Window, col 
sep=comma] {Part1_Exp2_2.dat};
\addplot[mark=triangle,black,mark options={scale=1.3},line width=1pt] table [x=EbNo, y=Haar, col 
sep=comma] {Part1_Exp2_2.dat};
\addplot[mark=o,black!20!red,line width=1pt] table [x=EbNo, y=NoNBI, col 
sep=comma] {Part1_Exp2_2.dat};
\end{semilogyaxis}
\end{tikzpicture}
\end{center}}
\caption{BER performance comparison of Haar transform and windowing based proposed sparse NBI reconstruction as a function of $E_b/N_0$ with $|\Tc|\%=25\%$ under grid offset.}
\label{fig:part1_exp2_2}
\end{figure}
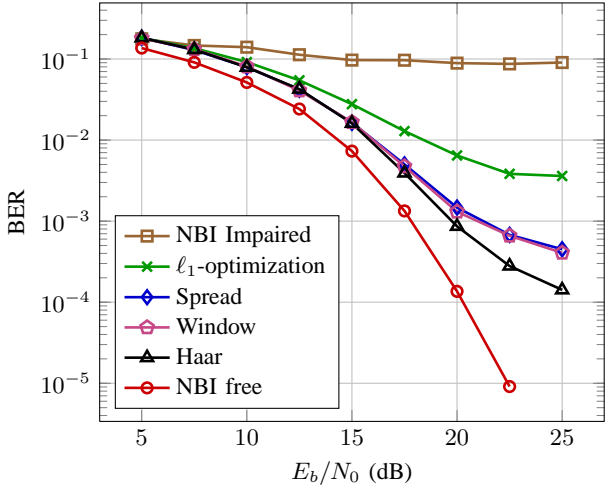
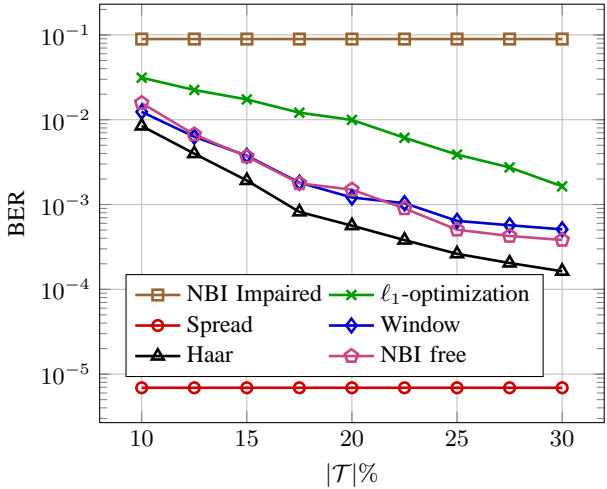
\begin{figure}[h!]\centering
{\small\begin{tikzpicture}
\begin{semilogyaxis}[
xlabel={$|\Tc|\%$},
ylabel={BER},
ylabel near ticks,
yminorticks,
xlabel near ticks,
width=0.45\textwidth,
legend cell align=left,
legend style={at={(0.88,0.355)}},
legend columns=2,
legend entries={NBI Impaired, $\ell_1$-optimization, Spread, Window, Haar, NBI free},
grid=major,
]
\addplot[mark=square,black!20!brown,line width=1pt] table [x=TS, y=NBI, col 
sep=comma] {Part1_Exp2_3.dat};
\addplot[mark=x,black!40!green,mark options={scale=1.3},line width=1pt] table [x=TS, y=l1, col 
sep=comma] {Part1_Exp2_3.dat};
\addplot[mark=o,black!20!red,line width=1pt] table [x=TS, y=NoNBI, col 
sep=comma] {Part1_Exp2_3.dat};
\addplot[mark=diamond,black!20!blue,mark options={scale=1.3},line width=1pt] table [x=TS, y=SABMP, col 
sep=comma] {Part1_Exp2_3.dat};
\addplot[mark=triangle,black,mark options={scale=1.3},line width=1pt] table [x=TS, y=Haar, col 
sep=comma] {Part1_Exp2_3.dat};
\addplot[mark=pentagon,black!20!magenta,mark options={scale=1.3},line width=1pt] table [x=TS, y=Window, col 
sep=comma] {Part1_Exp2_3.dat};
\end{semilogyaxis}
\end{tikzpicture}%
}
\caption{BER performance comparison of Haar transform and windowing based proposed sparse NBI reconstruction as a function of $|\Tc|$ with $E_b/N_0=22.5$dB under grid offset.}
\label{fig:part1_exp2_3}
\end{figure}
\subsubsection{Experiment 3: Noise-cancellation in ZF-FDE}
\begin{figure}[h!]\centering
{\small\begin{tikzpicture}
\begin{semilogyaxis}[
xlabel={$E_b/N_0$ (dB)},
ylabel={BER},
ylabel near ticks,
yminorticks,
xlabel near ticks,
xtick={5,10,15,20,25},
width=0.45\textwidth,
legend cell align=left,
legend pos=south west,
legend columns=1,
legend entries={ZF,ZF with Noise-cancellation,MMSE},
grid=major,
]
\addplot[mark=square,black!40!green,line width=1pt] table [x=EbNo, y=LS, col 
sep=comma] {Part1_Exp3_1.dat};
\addplot[mark=diamond,black!10!orange,mark options={scale=1.3},line width=1pt] table [x=EbNo, y=SABMP, col 
sep=comma] {Part1_Exp3_1.dat};
\addplot[mark=o,black!20!magenta,line width=1pt] table [x=EbNo, y=MMSE, col 
sep=comma] {Part1_Exp3_1.dat};
\end{semilogyaxis}
\end{tikzpicture}%
}
\caption{Noise-cancellation performance of the proposed sparse reconstruction scheme for ZF-FDE as a function of $E_b/N_0$ with $|\Tc|\%=25\%$.}
\label{fig:part1_exp3_1}
\end{figure}
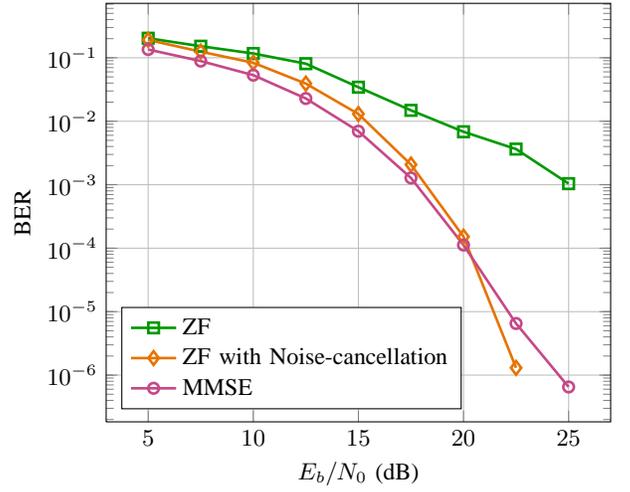

In this experiment, the noise-cancellation ability of the proposed sparse reconstruction scheme is examined under ZF-FDE. The sub-sampling rate is kept fixed at $1/4$ and the proposed noise-cancellation scheme is compared with the MMSE-FDE. Observe from the results of Fig.~\ref{fig:part1_exp3_1} that the proposed noise-cancellation scheme provides an error rate that is comparable to the MMSE-FDE.
\section{Reliable Carriers for Augmented NBI Recovery}\label{sec:ANBIR}
So far, we have completely relied on reserved tones to estimate the NBI. Now, to improve the spectral efficiency, we reduce the number of reserved tones and compensate for that with a data-aided approach. Note that for low SIR, a majority of data points in received NBI impaired signal are out of their correct decision regions (as demonstrated by the high BER of the (\emph{NBI impaired}) curves in Fig.~\ref{fig:part1_exp1_1},\ref{fig:part1_exp1_2},\ref{fig:part1_exp1_3},\ref{fig:part1_exp2_2},\ref{fig:part1_exp2_3}). However, after subtracting the NBI estimate from (\ref{eqestnbi}), we have
\begin{align}\label{eqestdnbi}
\hat{\xv}_u=\xv_u+\Em_u\underbrace{(\Ifd^\prime-\hat{\Ifd}^\prime)}_{\tilde\Ifd}=\xv_u+\dv_u,
\end{align}
where $\dv_u=\Em_u\tilde\Ifd^\prime$, and is henceforth called the \emph{residual NBI}. At this stage, it is reasonable to assume that the residual NBI is not strong and a good number of data points lie in their correct decision regions. These data points can be utilized to further reduce the residual NBI. Since the NBI has only few dominant elements, we only need a few data carriers to sense it. Therefore, we look for subset of the most reliable carriers, i.e., the carriers that have a high probability of falling in their correct decision regions\footnote{The severe impact of the NBI disallows the use of data-aided NBI recovery from the outset (to completely eliminate the requirement for reserved tones). Nonetheless, reliable carriers can be used in conjunction with reserved tones to improve the system spectral efficiency.}. There are two fundamental questions associated with the use of data-aided approach: i) how to find a subset of reliable data carriers and ii) once this subset is formed, how to use it in conjunction with the reserved tones to improve the reconstruction accuracy. We start by addressing the second question and later present a systematic procedure to select a subset of data points that is reliable.

Let us assume that a set of $|\Rc|$ reliable carriers indexed by $\Rc$ is available, where $\Rc\cap\Tc=\emptyset$. Now proceed by projecting (\ref{eqestnbi}) onto a binary selection matrix $\Sm_\Rc$ and obtain
\begin{align}\label{dataaidednbisys}
&\Sm_{\Rc}\hat{\xv}_u=\Sm_{\Rc}\xv_u+\Sm_{\Rc}\Em_u(\Ifd+\Zfd),\nonumber\\
\Longrightarrow&\underbrace{\Sm_{\Rc}\hat{\xv}_u-\Sm_{\Rc}\xv_u}_{\xv_{\Rc}^\prime}=\underbrace{\Sm_{\Rc}\Em_u}_{\Psim_{\Rc}}\underbrace{(\Ifd+\Zfd)}_{\Ifd^\prime}.
\end{align}
where $\Sm_\Rc$ is a binary selection matrix, with one entry equal to $1$ per row corresponding to the location of a reliable data point. This equation has the same form as (\ref{equndetsys}) and can be simply written as $\xv_\Rc^\prime=\Psim_\Rc\Ifd^\prime$. Note that, the unknown $\Ifd^\prime$ is identical to (\ref{equndetsys}) and the sensing matrix $\Psim_\Rc$ is similar to the sensing matrix $\Psim_{\Tc}$. Hence given the measurements $\xv_\Rc^\prime$ are available, we can use the set of equation (\ref{dataaidednbisys}) in conjunction with (\ref{equndetsys}) to find a better estimate of $\Ifd^\prime$. Concatenating (\ref{equndetsys}) and (\ref{dataaidednbisys}), we get
\begin{align}\label{eq:jointnullrel}
\underbrace{
\begin{bmatrix}
\xv_\Tc^\prime\\
\xv_\Rc^\prime
\end{bmatrix}}_{\xv^\prime}
=
\underbrace{
\begin{bmatrix}
\Psim_\Tc\\
\Psim_\Rc
\end{bmatrix}}_{\Psim}
\Ifd^\prime,
\Longrightarrow
\xv^\prime=\Psim\Ifd^\prime.
\end{align}
In comparison with (\ref{equndetsys}) (which had $|\Tc|$ equations), (\ref{eq:jointnullrel}) has $|\Tc|+|\Rc|$ equations, and hence the solution of (\ref{eq:jointnullrel}) is expected to provide an improved NBI estimate. However, to solve (\ref{eq:jointnullrel}), we require the set of measurements $\xv_\Rc^\prime=\Sm_{\Rc}\hat{\xv}_u-\Sm_{\Rc}\xv_u$. To obtain $\Sm_\Rc\xv_u$, we proceed by projecting (\ref{eqestdnbi}) on $\Sm_\Rc$, that yields
\begin{align}\label{eqrelcarriersfind}
\Sm_\Rc\hat{\xv}_u=\Sm_\Rc\xv_u+\Sm_\Rc\dv_u.
\end{align}
Under the assumption that the set $\Rc$ indexes the reliable carriers, the equality $\Sm_\Rc\lfloor\hat{\xv}_u\rfloor=\Sm_\Rc\xv_u$ holds in (\ref{eqrelcarriersfind}), where $\lfloor\cdot\rfloor$ represents a maximum-likelihood decision (i.e., it rounds its argument to the nearest QAM constellation point). Further, note that $\Sm_\Rc\hat{\xv}_u$ is accessible via (\ref{dataaidednbisys}), and hence, both components required for the evaluation of $\xv_\Rc^\prime$ are available. Now, (\ref{eq:jointnullrel}) can be solved for $\Ifd^\prime$ using SABMP.

Now let us go back to the first question i.e., determining the subset of reliable carriers $\Rc$. To obtain this index set note that, in (\ref{eqestdnbi}), we expect the following: for some sub-carriers, the residual NBI $d(i)$ is strong enough to take $x(i)$ out of its correct decision region i.e., $\lfloor\hat{x}(i)\rfloor\neq x(i)$, while for others with a milder residual NBI, we expect to have $\lfloor\hat{x}(i)\rfloor=x(i)$. The subset of data carries which satisfy $\lfloor\hat{x}(i)\rfloor=x(i)$ are the reliable carriers and fortunately constitute a major part of all data sub-carriers (after initial NBI compensation). To select this subset, we note that the major source of perturbation is the residual NBI distortion, especially for high signal-to-noise ratio. Hence, we can write the reliability function for the $i{\rm th}$ sub-carrier in terms of $d(i)$ as
\begin{align}\label{eqreliability}
\mathfrak{R(i)}=\dfrac{p(d(i)=\hat{x}(i)-\lfloor\hat{x}(i)\rfloor)}{\sum_{q=0,\Ac(q)\neq\lfloor\hat{x}(i)\rfloor}^{Q-1}p(d(i)=\hat{x}(i)-\Ac(q))},
\end{align}
where $p(\cdot)$ represents the pdf of $d$, which is assumed to be zero mean Gaussian with variance $\sigma_d^2$ (see \cite{al2012pilotless} for details)\footnote{While evaluating $\mathfrak{R}$ in (\ref{eqreliability}), the the residual distortion $d$ is assumed normal. Assuming a known prior is necessary in this case in order to obtain a closed form expression for $\mathfrak{R}$. Further, numerical results demonstrate satisfactory performance of the utilized reliability metric under the assumption that $d$ is normal.}. Note that once the covariance of the residual NBI $\tilde\Ifd^\prime$ is available, the covariance matrix of the distortion $\dv$ (and hence $\sigma_d^2$) can be obtained. The procedure of obtaining the covariance of the estimation error term $\tilde\Ifd$ is outlined in Appendix~\ref{AppB}. In (\ref{eqreliability}), the numerator is the probability that $d(i)$ does not take $x(i)$ beyond its correct decision region and the denominator sums the probabilities of all possible incorrect decisions that $d(i)$ can cause. After obtaining the reliability $\mathfrak{R}(i)$ for each carrier $i$, we pick the $|\Rc|$ sub-carriers with highest reliability values and index them using $\Rc$. This index set is used in the previously discussed manner to reconstruct the unknown clipping vector. Further, it is observed that the set $\Rc$ constructed using the reliability metric (\ref{eqreliability}) indexes non-uniformly placed tones and hence is fitting for CS based sparse recovery.

The concept of using reliable data carriers is not new, (see e.g., \cite{ali2013compressed,owodunni2014compressed}). However, the reliability criteria employed (by some of the autors in \cite{ali2013compressed,owodunni2014compressed}) is simplistic and relies solely on the relative distance of the received constellation point from its neighbours to determine the confidence level. In order to explain the distance based approach, we consider as a motivating example the constellation shown in Fig.~\ref{da_fig}. Here $\hat x_1$ and $\hat x_2$ are two equalized data samples which are equidistant from the closest constellation point $x$. However, in spite of being equidistant from $x$, $\hat x_1$ and $\hat x_2$ have different reliability values. This is because the distances of these two points from their respective next nearest neighbours are different. Specifically, note that $x_a$ is next nearest neighbour of $\hat x_1$ and $x_b$ is next nearest neighbour of $\hat x_2$ respectively. Given that $\hat x_1$ and $\hat x_2$ are equidistant from $x$, $\hat x_2$ is considered more reliable than $\hat x_1$ since in relative terms, we have
\begin{equation}
\dfrac{|\hat x_2-x|}{|\hat x_2- x_b|}<\dfrac{|\hat x_1-x|}{|\hat x_1-x_a|}
\end{equation}
This motivated the following reliability metric $\mathfrak{R}(i)$,
\begin{equation}
\mathfrak{R}(i)=-\log\left(\dfrac{|\hat x-\lfloor\hat x\rfloor|}{|\hat x-\lfloor\hat x\rfloor_{NN}|}\right)
\end{equation}
where as defined before, $\lfloor\hat x\rfloor$ denotes rounding to the nearest constellation point while $\lfloor\hat x\rfloor_{NN}$ denotes rounding to the next nearest constellation point. Thus, after calculating the reliability of all $N$ carriers, the reliabilities were sorted in descending order $\mathfrak{R}(i_1)\geq\mathfrak{R}(i_2)\geq\cdots\geq\mathfrak{R}(i_N)$ and $|\Rc|$ carriers with the highest reliability were chosen.
\usetikzlibrary{shadows,fadings,patterns,snakes,positioning,decorations.markings,backgrounds}
\begin{figure}[h!]\centering
\begin{tikzpicture}[framed,scale=0.8]
\node (a) at (0,0) {\large $\pmb\times$};
\node at (0,-5) {\large $\pmb\times$};
\node at (-5,0) {\large $\pmb\times$};
\node at (-5,-5) {\large $\pmb\times$};

\node at (-5+0.5,-5+0.5) {\Large $x_c$};
\node at (-5+5.5,-5+0.5) {\Large $x_b$};
\node at (-5+0.5,-5+5.5) {\Large $x_a$};
\node at (-5+5.5,-5+5.5) {\Large $x$};

\draw [black,dashed,thick] (0,0) circle [radius=1.7];

\filldraw (-0.8,-1.5) circle (3pt);
\filldraw (-1.7,0) circle (3pt);

\draw [thick] (-5,0)--(-1.7,0);
\draw [dashed,thick] (-1.7,0)--(0,0);
\draw [thick] (0,-5)--(-0.8,-1.5);
\draw [dashed,thick] (0,0)--(-0.8,-1.5);

\node at (-1.7-0.5,0+0.5) {\Large $\hat x_1$};
\node at (-0.8-0.5,-1.5-0.5) {\Large $\hat x_2$};

\end{tikzpicture}
\caption{Geometrical representation of adopted reliability criteria.}
\label{da_fig}
\end{figure}
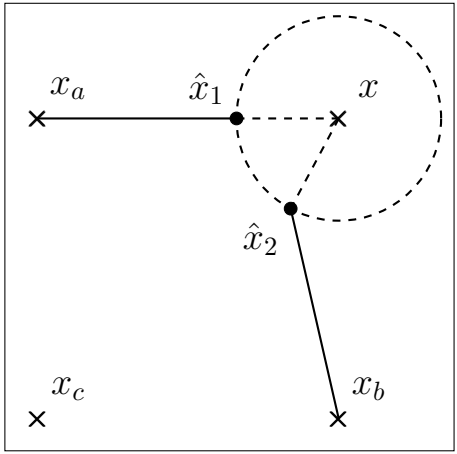
In comparison to the distance based reliability metric, the reliability function utilized in this work (given in (\ref{eqreliability})) makes use of the additional information about residual NBI (i.e., $\sigma_d^2$) and is rigorous towards analyzing the reliability of the data points. Note also that the advantage of spectral efficiency expected by the use of data-aided approach comes at the expense of increased computational complexity. As the data-aided reconstruction is a two stage process, its complexity is roughly twice the computational complexity of the one stage (reserved tones only) reconstruction.

The selection of $|\Tc|$ and $|\Rc|$ is also critical for NBI mitigation. From sparse signal reconstruction point of view, it is desirable to increase the measurements, however, other limitations render it infeasible to choose an arbitrarily large $|\Tc|$ and $|\Rc|$. The choice of $|\Tc|$ is mainly dictated by the desired data rate and the number of NBI sources that the system may experience. The selection of $|\Rc|$ is dependent on the conflicting interests associated with increasing or decreasing the number of utilized reliable tones. Note that, though a larger $|\Rc|$ promises improved estimation accuracy but at the same time the risk of feeding erroneous information to the reconstruction algorithm is also increased.
\subsection{Simulation Results}
In this subsection, the effectiveness of the proposed data-aided reconstruction scheme is numerically assessed. The general simulation setup (i.e., sub-carriers, users, channel length, SIR, QAM order etc.) is same as in the simulation subsection \ref{ref:simres1}. However, changes whenever made are highlighted in the description of experiments. In the first experiment, the BER of the proposed data-aided reconstruction scheme is compared with \emph{reserved tones only} reconstruction. Whereas the second experiment is carried to demonstrate the relationship between $|\Tc|$ and $|\Rc|$.
\subsubsection{Experiment 1: Data-aided CS for Spectral Efficiency}
We start by choosing $|\Tc|=32$ and use $|\Rc|=|\Tc|=32$ for data-aided NBI reconstruction. This way, though the sub-sampling rate is still ($\frac{|\Tc|+|\Rc|}{N}=\frac{64}{256}=\frac{1}{4}$), the reserved tones rate ($\frac{|\Tc|}{N}=\frac{32}{256}=\frac{1}{8}$) is cut into half. We start with the general grid offset case where a maximum of four NBI sources per symbol are assumed. The recovery results are generated by using Haar transform based sparsification. From the results (shown in Fig. \ref{fig:Part2_Exp1_1}), it is evident that the signal reconstruction accuracy considerably improves by using reliable tones in conjunction with reserved tones (\emph{Augmented-Recovered}). Further, the proposed data-aided scheme is also tested for the optimistic case of no grid offset and the results are shown in Fig.~\ref{fig:Part2_Exp1_2}. The findings for perfect alignment are inline with the observations for the grid offset case and depict the advantage of using reliable tones.
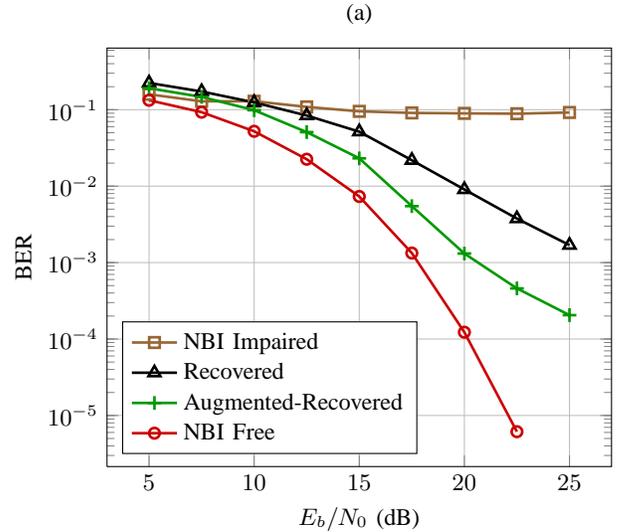
\begin{figure}[h!]\centering
{\small\begin{tikzpicture}[baseline]
\begin{semilogyaxis}[
title={(a)},
name=offset,
xlabel={$E_b/N_0$ (dB)},
ylabel={BER},
ylabel near ticks,
yminorticks,
xlabel near ticks,
xtick={5,10,15,20,25},
width=0.45\textwidth,
legend cell align=left,
legend pos=south west,
legend columns=1,
legend entries={NBI Impaired, Recovered, Augmented-Recovered,NBI Free},
grid=major,
]
\addplot[mark=square,black!20!brown,line width=1pt] table [x=EbNo, y=NBI, col 
sep=comma] {Part2_Exp1_1.dat};
\addplot[mark=triangle,black,mark options={scale=1.3},line width=1pt] table [x=EbNo, y=SABMP, col 
sep=comma] {Part2_Exp1_1.dat};
\addplot[mark=+,black!40!green,mark options={scale=1.3},line width=1pt] table [x=EbNo, y=SABMPA, col 
sep=comma] {Part2_Exp1_1.dat};
\addplot[mark=o,black!20!red,line width=1pt] table [x=EbNo, y=NoNBI, col 
sep=comma] {Part2_Exp1_1.dat};
\end{semilogyaxis}
\end{tikzpicture}}
\caption{BER performance of the proposed data-aided sparse recovery scheme as a function of $E_b/N_0$ with $|\Tc|\%=|\Rc|\%=12.5\%$ under grid offset and Haar transform based sparsity restoration.}
\label{fig:Part2_Exp1_1}
\end{figure}
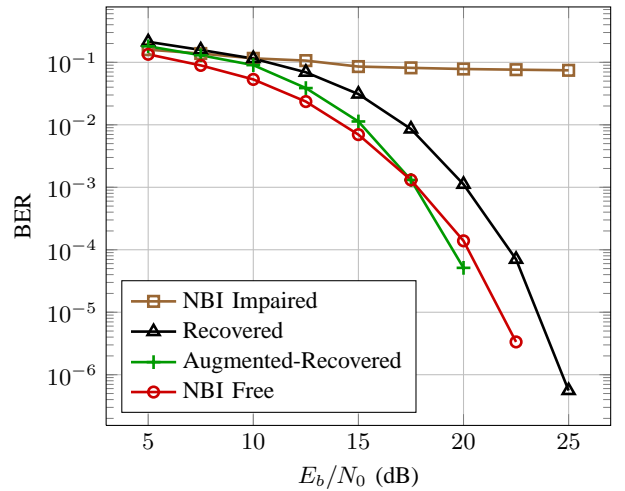
\begin{figure}[h!]\centering
{\small\begin{tikzpicture}
\begin{semilogyaxis}[
xlabel={$E_b/N_0$ (dB)},
ylabel={BER},
ylabel near ticks,
yminorticks,
xlabel near ticks,
xtick={5,10,15,20,25},
width=0.45\textwidth,
legend cell align=left,
legend pos=south west,
legend columns=1,
legend entries={NBI Impaired, Recovered, Augmented-Recovered,NBI Free},
grid=major,
]
\addplot[mark=square,black!20!brown,line width=1pt] table [x=EbNo, y=NBI, col 
sep=comma] {Part2_Exp1_2.dat};
\addplot[mark=triangle,black,mark options={scale=1.3},line width=1pt] table [x=EbNo, y=SABMP, col 
sep=comma] {Part2_Exp1_2.dat};
\addplot[mark=+,black!40!green,mark options={scale=1.3},line width=1pt] table [x=EbNo, y=SABMPA, col 
sep=comma] {Part2_Exp1_2.dat};
\addplot[mark=o,black!20!red,line width=1pt] table [x=EbNo, y=NoNBI, col 
sep=comma] {Part2_Exp1_2.dat};
\end{semilogyaxis}
\end{tikzpicture}%
}
\caption{BER performance of the proposed data-aided sparse recovery scheme as a function of $E_b/N_0$ with $|\Tc|\%=|\Rc|\%=12.5\%$ for the optimistic case of perfect grid alignment.}
\label{fig:Part2_Exp1_2}
\end{figure}

\subsubsection{Experiment 2: Choosing $|\Tc|$ and $|\Rc|$}
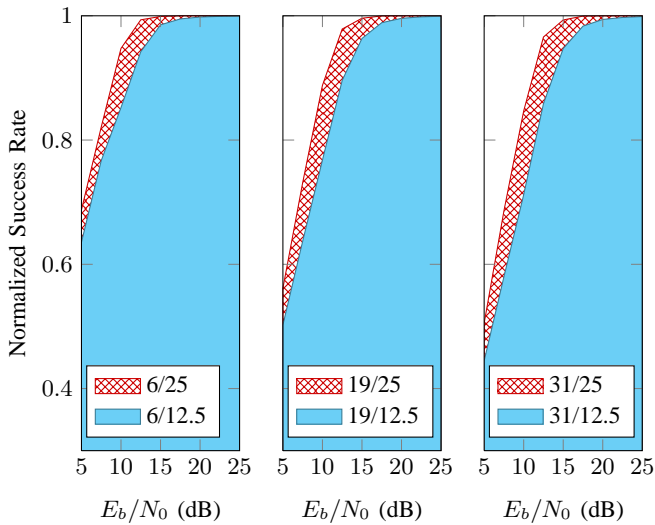
\begin{figure}[h!]\centering
{\small\begin{tikzpicture}[baseline]
\begin{axis}[area style,
xmin=5,xmax=25,
ymin=0.3,ymax=1,
xlabel={$E_b/N_0$ (dB)},
ylabel={Normalized Success Rate},
ylabel near ticks,
ytick={0.4,0.6,0.8,1},
height=0.4\textwidth,
width=0.2\textwidth,
legend cell align=left,
legend entries={6/25,6/12.5},
legend pos=south west,
legend columns=1,
]
\addplot[pattern=crosshatch,pattern color=black!20!red,draw=black!20!red,
samples=200] table [x=EbNo, y=1664, col 
sep=comma] {Part2_Exp1_3.dat} \closedcycle;
\addplot[draw=black!50!cyan,fill=white!50!cyan] table [x=EbNo, y=1632, col 
sep=comma] {Part2_Exp1_3.dat} \closedcycle;
\end{axis}
\end{tikzpicture}
\hspace{-3pt}
\begin{tikzpicture}[baseline]
\begin{axis}[area style,
xmin=5,xmax=25,
ymin=0.3,ymax=1,
xlabel={$E_b/N_0$ (dB)},
ytick={0.4,0.6,0.8,1},
yticklabels={,,},
height=0.4\textwidth,
width=0.2\textwidth,
legend cell align=left,
legend entries={19/25,19/12.5},
legend pos=south west,
legend columns=1, 
]
\addplot[pattern=crosshatch,pattern color=black!20!red,draw=black!20!red,
samples=200] table [x=EbNo, y=4864, col 
sep=comma] {Part2_Exp1_3.dat} \closedcycle;
\addplot[draw=black!50!cyan,fill=white!50!cyan] table [x=EbNo, y=4832, col 
sep=comma] {Part2_Exp1_3.dat} \closedcycle;
\end{axis}
\end{tikzpicture}
\hspace{-3pt}
\begin{tikzpicture}[baseline]
\begin{axis}[area style,
xmin=5,xmax=25,
ymin=0.3,ymax=1,
xlabel={$E_b/N_0$ (dB)},
ytick={0.4,0.6,0.8,1},
yticklabels={,,},
height=0.4\textwidth,
width=0.2\textwidth,
legend cell align=left,
legend entries={31/25,31/12.5},
legend pos=south west,
legend columns=1,
]
\addplot[pattern=crosshatch,pattern color=black!20!red,draw=black!20!red,
samples=200] table [x=EbNo, y=8064, col 
sep=comma] {Part2_Exp1_3.dat} \closedcycle;
\addplot[draw=black!50!cyan,fill=white!50!cyan] table [x=EbNo, y=8032, col 
sep=comma] {Part2_Exp1_3.dat} \closedcycle;
\end{axis}
\end{tikzpicture}}
\caption{Normalized success rate for the reliability criteria as a function of reserved tones. The $|\Rc|\%/|\Tc|\%$ format represents the ratio of the percentage of reliable carriers picked $|\Rc|\%$ to the percentage of reserved tones $|\Tc|\%$.}
\label{fig:Part2_Exp1_3}
\end{figure}

This experiment is carried to demonstrate a direct relationship between the number of reliable tones available and the reserved tones. Note that, not all $|\Rc|$ carriers chosen using the proposed reliability criteria are guaranteed to fall in their correct decision region. Hence, we introduce a metric called normalized success rate defined as $|\{i: \lfloor\hat{x}(i)\rfloor=x(i),i\in \Rc\}|/|\Rc|$. This metric is used as a measure that shows that among $|\Rc|$ carriers chosen, how many carriers actually fall in their correct decision regions. We compare and plot normalized success rate as a function of $E_b/N_0$, for several value of the ratio $|\Rc|\%/|\Tc|\%$. The ratio $|\Rc|\%/|\Tc|\%$, depicts the percentage of reliable carriers $|\Rc|\%$ chosen, when the reserved tones were $|\Tc|\%$. In this experiment, the maximum number of active NBI sources in any symbol is limited to four, with varying locations and grid offset per symbol. The results of this experiment are shown in Fig.~\ref{fig:Part2_Exp1_3}. The crosshatched region shows the success rate for $25\%$ reserved tones, whereas, the solid region shows the success rate for $12.5\%$ reserved tones. It is observed that the number of correct decisions (measured as normalized success rate) based on the utilized reliability criteria will significantly increase, if we choose more reserved tones. The results are obviously expected, as more reserved tones result in better NBI reconstruction in the first stage and hence lower residual NBI, yielding more reliable carriers.
\section{Multiple Antenna Base-station}\label{sec:MABS}
In this section, we consider a BS equipped with multiple receiving antennas i.e., a SIMO setup. In this case, each antenna element will be subjected to the same transmitted signal impaired by the NBI sources. For a SIMO receiver, although it will be too restrictive to assume the same NBI signal across all antenna elements, it is practical to expect the NBI signals to share a common support (i.e., to consider the received NBI signals \emph{jointly-sparse}). However, the values of the active elements can vary across the antennas. Just as the transmitted data will arrive at different antennas through different channels, the antennas will experience the same NBI possibly through different channels. In other words, the NBI vectors that each antenna needs to mitigate share a common support but possibly different values at the active elements. For $T$ antenna system, we will have $T$ equations of the form (\ref{eq:jointnullrel}), one belonging to each antenna element. Thus we have
\begin{align}
\xv_1^\prime=&\Psim_1\Ifd_1^\prime\nonumber\\
\xv_2^\prime=&\Psim_2\Ifd_2^\prime\nonumber\\
&\vdots\nonumber\\
\xv_T^\prime=&\Psim_T\Ifd_T^\prime\nonumber
\end{align}
where $\xv_t^\prime$, $\Psim_t$ and $\Ifd_t^\prime$ represent the measurement vector, sensing matrix and the unknown NBI signal belonging to the antenna number $t$. This gives rise to the multiple measurement vector (MMV) problem, where the antennas collaborate in figuring out the common support (of $\Ifd_t^\prime$s) and subsequently each antenna individually estimates the NBI it sees at this support. The SABMP algorithm that has been used so far can be extended to the MMV setup (i.e., MMV-SABMP) \cite{al2013distribution}. The cornerstone of MMV-SABMP scheme is to find the support of the unknown signal collaboratively (based on all measurement vectors) and later estimate the amplitudes of the active elements individually for each unknown vector. Here it is worth highlighting that the joint-sparse NBI estimation can be casted as a block sparse problem and can be solved using either SABMP or $\ell_1$-optimization. However, the block sparse signal estimation requires processing at a central location, that has access to all the measurement vectors. In certain applications, communicating the measurements to a central location can impose excessive communication burden. Whereas, the MMV-SABMP based joint-estimation can be carried in a distributed manner, with minimum communication overhead between antennas. Note that the computational load of the MMV-SABMP algorithm is $\Oc(|\Tc|sN^2T^2)$. However, if NBI reconstruction on all antennas is performed in parallel, the computational burden becomes linear in $T$ i.e., $\Oc(|\Tc|sN^2T)$.
\subsection{Simulation Results}
In this subsection, the performance of MMV based NBI reconstruction is compared with single measurement vector (SMV) based reconstruction. The number of receiver antennas is assumed to be $2$ and received signals are combined using maximal ratio combining (MRC). Other simulation parameters are kept consistent with the previously explained setup.
\subsubsection{SMV vs MMV Reconstruction for Jointly-Sparse NBI}
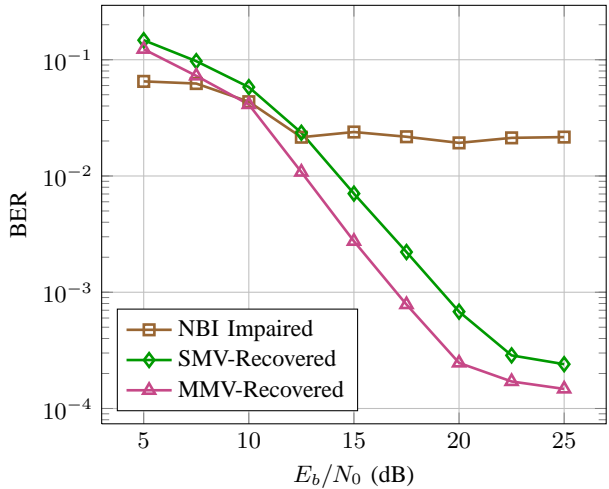
\begin{figure}[h!]\centering
{\small\begin{tikzpicture}
\begin{semilogyaxis}[
xlabel={$E_b/N_0$ (dB)},
ylabel={BER},
ylabel near ticks,
yminorticks,
xlabel near ticks,
xtick={5,10,15,20,25},
width=0.45\textwidth,
legend cell align=left,
legend pos=south west,
legend columns=1,
legend entries={NBI Impaired, SMV-Recovered, MMV-Recovered},
grid=major,
]
\addplot[mark=square,black!20!brown,line width=1pt] table [x=EbNo, y=NBI, col 
sep=comma] {Part3_Exp1_1.dat};
\addplot[mark=diamond,black!40!green,mark options={scale=1.3},line width=1pt] table [x=EbNo, y=Ind, col 
sep=comma] {Part3_Exp1_1.dat};
\addplot[mark=triangle,black!20!magenta,mark options={scale=1.3},line width=1pt] table [x=EbNo, y=Jnt, col 
sep=comma] {Part3_Exp1_1.dat};
\end{semilogyaxis}
\end{tikzpicture}%
}
\caption{BER performance of the proposed MMV based reconstruction of the jointly-sparse NBI signal as a function of $E_b/N_0$ for $|\Tc|\%=12.5\%$ under grid offset.}
\label{fig:Part3_Exp1_1}
\end{figure}

For simulation, the number of reserved tones is kept fixed at $|\Tc|=32$ and hence, the sub-sampling rate is $\frac{|\Tc|}{N}=\frac{32}{256}=\frac{1}{8}$. The maximum number of active NBI sources is four and the practical case of grid offset is considered. The BER performance of MMV and SMV reconstruction is compared for varying $E_b/N_0$ and results are presented in Fig.~\ref{fig:Part3_Exp1_1}. It can be observed that the MMV reconstruction improves the BER over SMV reconstruction throughout the range of interest.
\section{Conclusion and Future Work}\label{sec:Conc}
In this paper, we have developed a framework for NBI mitigation in SC-FDMA systems. The proposed approach utilizes the sparsity (or compressibility) of the NBI signal and makes use of a Bayesian algorithm (i.e., SABMP) for the estimation and cancellation of NBI. The SABMP algorithm has several advantages over other sparse reconstruction algorithms, including i) low estimation error, ii) low complexity, iii) ambivalence to the distribution of the sparse vector and iv) availability of estimate error metric. In addition, to maximize the spectral efficiency of the proposed scheme, a data-aided approach was presented. In this regard, a systematic procedure was outlined to select the most reliable carriers and their use in conjunction with reserved tones. The proposed reconstruction scheme was also extended to the practical case of multiple receiver antennas where the distributed version of the SABMP i.e, MMV-SABMP is used. This distributed version is essential in reducing the complexity of the MMV-SABMP algorithm and makes it suitable for NBI mitigation in multicell scenarios.

Studying the bandwidth efficiency of the proposed scheme remains for future work. This is an important problem as it can help determine the minimum number of measurements required (i.e., $|\Tc|_{min}$) to reconstruct an NBI vector of length $N$, given that it has $sN$ active elements.
\appendix
\subsection{A Brief Description of SABMP}\label{AppA}
Let us estimate the sparse vector $\Ifd\in\mathbb{C}^{N}$, from the the observations vector $\xv\in\mathbb{C}^{M}$ related by the linear regression model, $\xv = \Psim \Ifd + \zv$. Here, $\Psim\in\mathbb{C}^{M\times N}$ and $\zv\sim\Nc\Cc(\zerov, \sigma_z^2\Id)$. The SABMP algorithm pursues an MMSE estimate of $\Ifd$ given $\xv$ as follows,
\begin{equation}\label{eq:app:xmmse}
\hat{\Ifd}_{\rm{MMSE}} \triangleq \mathbb{E}[\Ifd|\xv] = \sum_{\Sc} p(\Sc|\xv)\mathbb{E}[\Ifd|\xv,\Sc].
\end{equation}
Here the sum is executed over all possible $2^N$ support sets $\Sc$ of $\Ifd$. However, computing this sum is a challenging task when $(N)$ is large, because the number of possible support sets can be extremely large and the computational complexity will become unrealistic. To have a computationally feasible solution, this sum can be approximated by considering only those support sets which include the most significant taps with high probability. These few support sets correspond to the sets with significant posteriors $p(\Sc|\xv)$. Let $\Sc_d$ be the set of supports for which the posteriors are significant. Hence, (\ref{eq:app:xmmse}) can be approximated to find the approximate estimate $\hat{\Ifd}_{\rm AMMSE}$ as follows,
\begin{equation}\label{eq:app:xammse}
\hat{\Ifd}_{\rm{AMMSE}} = \sum_{\Sc\in\Sc_d} p(\Sc|\xv)\mathbb{E}[\Ifd|\xv,\Sc].
\end{equation}
We could determine $\Sc_d$ and $\hat{\Ifd}_{\rm AMMSE}$ in a greedy manner using the dominant support selection metric defined as the log posterior,
\begin{align}\label{eq:app:dssm1}
\nu(\Sc) &\triangleq \ln p(\Sc |\xv) = \ln p(\xv|\Sc) p(\Sc).
\end{align}

The greedy algorithm of SABMP starts by first finding the best support of size 1. This requires evaluating $\nu(\Sc)$ for $\Sc=\{ 1 \}, \dots, \{ N \}$, i.e., a total of $\binom{N}{1}$ search points. Let $\Sc_1 = \{ \alpha_1 \}$ be the optimal support. Now, the optimal support of size 2 is found. Ideally, this involves a search over a space of size $\binom{N}{2}$. To reduce the search space, however, the greedy approach looks for the tap location $\alpha_2 \neq \alpha_1$ such that $\Sc_2=\{ \alpha_1, \alpha_2 \}$ maximizes $\nu(\Sc_2)$. This involves $\binom{N-1}{1}$ search points (as opposed to the optimal search over $\binom{N}{2}$ points). The process continues in this manner by forming $\Sc_3 = \{ \alpha_1, \alpha_2, \alpha_3 \}$ and so on. Therefore, $\Sc_d$, the set of dominant support sets is composed of support sets that are incremental in nature as shown below,\footnote{In (\ref{eq:app:Sd}), $T_{\mathrm{max}}$ refers to the maximum number of non-zero elements in the sparse $\hv$. $T_{\mathrm{max}}$ is selected to be slightly larger than the expected number of active taps in the estimated sparse signal using the de Moivre-Laplace theorem. For details about the algorithm, readers are directed to the reference \cite{masood2013sparse}.}
\begin{align}\label{eq:app:Sd}
\Sc_d &= \left\{ \Sc_1, \Sc_2, \dots, \Sc_{T_{\mathrm{max}}}  \right\},\nonumber\\
\Sc_d &= \left\{  \{ \alpha_1\}, \{ \alpha_1, \alpha_2\}, \{ \alpha_1, \alpha_2, \alpha_3\}, \dots, \{ \alpha_1, \alpha_2, \dots, \alpha_{T_{\mathrm{max}}}\} \right\},
\end{align}

The development of the SABMP algorithm in \cite{masood2013sparse} assumes that entries of $\Ifd$ are activated with equal probability $\lambda$ (i.e., i.i.d. Bernoulli with probability $\lambda$). However, if some elements are more probable than others (based on the available information), it is desirable to assign those entries a higher probability. This requires us to assume a non-i.i.d. Bernoulli behavior for the unknown sparse vector \cite{panahi2011fast}, and therefore the prior is given as follows
\begin{align}\label{eq:app:pS}
p(\Sc) &=  \prod_{i\in \Sc} \lambda_i \prod_{j\in\{1,\dots,N\}\backslash \Sc } (1-\lambda_j),
\end{align}
where, $\lambda_i$ is the probability by which the $i$th element of $\Ifd$ is active. Moreover, the likelihood is approximated as follows,
\begin{align}\label{eq:app:pyS}
p(\xv|\Sc) &=  \exp \left( -\frac{1}{2\sigma_z^2} \left\|\Pm_\Sc^{\bot} \xv\right\|_2^2 \right),
\end{align}
where, $\Pm_\Sc^\bot =\Id - \Pm_\Sc  =\Id - \Psim_\Sc\left( \Psim_\Sc^\herm \Psim_\Sc\right)^{-1} \Psim_\Sc^\herm$ is the projection matrix and $\Psim_\Sc$ is formed by selecting the columns of $\Psim$ corresponding to the support $\Sc$. Further, given that $M > |\Sc|$ (a necessary condition for CS), the matrix $\Psim_\Sc^\herm\Psim_\Sc$ is guaranteed to be well conditioned. Substituting (\ref{eq:app:pS}) and (\ref{eq:app:pyS}) in (\ref{eq:app:dssm1}) yields
\begin{align}\label{eq:app:dssm2}
\nu(\Sc) \triangleq \ln p(\Sc | \xv) &= ( -\frac{1}{2\sigma_z^2}) \left\|\Pm_\Sc^{\bot} \xv\right\|_2^2 + \sum_{i\in \Sc}\ln \lambda_i \nonumber\\
&+ \sum_{j \in \{1,\cdots,N\}\backslash\Sc}\ln(1-\lambda_j)
\end{align}
Now the only term that is left to be evaluated in (\ref{eq:app:xammse}) is  $\mathbb{E}[\Ifd|\xv,\Sc]$. Note that it is difficult or even impossible to evaluate this quantity because the distribution of the active taps of $\Ifd$ is unknown. Therefore, we replace it by the best linear unbiased (BLUE) estimate as follows
\begin{align}\label{eq:BLUE}
\mathbb{E}[\Ifd|\xv,\Sc] \leftarrow \left( \Psim_\Sc^\herm \Psim_\Sc\right)^{-1} \Psim_\Sc^\herm \xv.
\end{align}
This provides us all the required quantities to evaluate $\hat{\Ifd}_{\rm AMMSE}$. Note that all parameters including $\sigma_z^2, \boldsymbol{\lambda}=\{\lambda_i\}_{i=1}^N$ and the possible size of support $T_{\rm max}$ need not be known and are estimated by the algorithm. The SABMP algorithm is summarized in Table~\ref{alg:greedy}.
\begin{table}
\begin{algorithmic}[1]
\Procedure{Greedy} {$\Am, \yv, \boldsymbol{\lambda}, \sigma_z^2, T_{\rm max}$} %\Comment{The g.c.d. of a and b}
\State \textbf{initialize} $J \gets \{1, 2, \hdots, N\},\, i \gets 1$
\State \textbf{initialize empty sets} $\Sc_{max},\, \Sc_d,\, p(\Sc_d|\xv),\, \mathbb{E}[\Ifd|\xv,\Sc_d]$
\State $J_i \gets J$
\While{$i \le T_{\rm max}$} %\Comment{We have the answer if r is 0}
\State $\Omega \gets  \{ \Sc_{max} \cup \{ \alpha_1 \},
 \Sc_{max} \cup \{ \alpha_2 \},
\cdots,
 \Sc_{max} \cup \{ \alpha_{|J_i|} \} \mid \alpha_k \in J_i\}$
\State \textbf{compute }$\{ \nu(\Sc_k) \mid \Sc_k \in \Omega \}$
\State \textbf{find} $\Sc_\star \in \Omega$ \textbf{such that} $\nu(\Sc_\star) \ge \max_j \nu(\Sc_j)$
\State $\Sc_d \gets \{\Sc_d, \Sc_\star\}$
\State \textbf{compute} $p(\Sc_\star|\xv), \mathbb{E}[\Ifd|\xv,\Sc_\star]$
\State $p(\Sc_d|\xv) \gets \{p(\Sc_d|\xv), p(\Sc_\star|\xv)\}$
\State $\mathbb{E}[\Ifd|\xv,\Sc_d] \gets \{\mathbb{E}[\Ifd|\xv,\Sc_d] , \mathbb{E}[\Ifd|\xv,\Sc_\star\}$
\State $\Sc_{max}\gets \Sc_\star$
\State $J_{i+1} \gets N~\backslash~\Sc_\star$
\State $i \gets i+1$
\EndWhile\label{euclidendwhile}
\State\label{alg:greedy:returnline} \textbf{return} $\Sc_d, p(\Sc_d|\xv), \mathbb{E}[\Ifd|\xv,\Sc_d]$
\EndProcedure
\end{algorithmic}
\caption{Support Agnostic Bayesian Matching Pursuit Algorithm (SABMP)}\label{alg:greedy}
\end{table}

\subsection{Error Covariance and Estimation Error}\label{AppB}
Let,
\begin{align}\label{eq:channelerror}
\tilde{\Ifd} = \hat{\Ifd}_{\rm AMMSE} - \Ifd
\end{align}
be the error vector and  $\Rm_{\tilde{\boldsymbol{\Ic}}} \triangleq {\rm cov}[\tilde{\boldsymbol{\Ifd}}|\xv]$ where $\rm cov$ represents the covariance. The trace of $\Rm_{\tilde{\boldsymbol{\Ic}}}$ i.e., $\textrm{Tr}[\Rm_{\tilde{\boldsymbol{\Ic}}}]$ gives the MMSE estimation error. In order to evaluate $\Rm_{\tilde{\boldsymbol{\Ic}}}$, let us define the error vector $\tilde{\Ifd}_\Sc = \hat{\Ifd}_\Sc - \Ifd$ for a given support $\Sc$, where $\hat{\Ifd}_\Sc = \mathbb{E}[\Ifd|\xv,\Sc]$. Let the corresponding error covariance matrix be $\Rm_{\tilde{\boldsymbol{\Ic}}|\Sc} \triangleq {\rm cov}[\tilde{\Ifd}|\xv,\Sc]$. Then $\Rm_{\tilde{\boldsymbol{\Ic}}}$ could be expressed in terms of $\Rm_{\tilde{\boldsymbol{\Ic}}|\Sc}$ by summing it over the dominant support set $\Sc_d$ as follows
\begin{align}\label{eq:Rh}
\Rm_{\tilde{\boldsymbol{\Ic}}} &= \sum_{\Sc\in\Sc_d} p(\Sc|\xv) \,\, \Rm_{\tilde{\boldsymbol{\Ic}}|\Sc}.
\end{align}
Since we replace $\mathbb{E}[\Ifd|\xv,\Sc]$ with a BLUE estimate, the conditional error covariance matrix will be $\Rm_{\tilde{\boldsymbol{\Ic}}|\Sc} = (\Psim_\Sc^\herm \Cm^{-1} \Psim_\Sc)^{-1}$ \cite{poor1994introduction} (where $\Cm = \sigma_z^2 \Id$ is the noise covariance matrix). Combining this fact with (\ref{eq:Rh}) we have,
\begin{align}\label{eq:error_covariance_matrix}
\Rm_{\tilde{\boldsymbol{\Ic}}}\! &=\! \sigma_z^2\sum_{\Sc\in\Sc_d} p(\Sc|\xv) \,\, (\Psim_\Sc^\herm \Psim_\Sc)^{-1}.
\end{align}
Note that the calculation of covariance matrix involves a matrix inversion term which is a computationally expensive task. However, we would like to highlight that these inverses are available as part of intermediate calculations in the SABMP algorithm and hence do not pose any additional burden. Although simple to compute, the error covariance matrix and the estimation error play a vital role in the development of the data-aided approach presented in Sec. \ref{sec:ANBIR}. Further, it is worth highlighting that such a calculation of the error covariance is not possible for $\ell_1$-optimization based sparse signal recovery.
\bibliographystyle{IEEEtran}
\bibliography{SCFDMA_NBI}
\end{document}